\documentclass[twocolumn, 10pt, aps, prl, superscriptaddress]{revtex4-1}

\usepackage{graphicx}
\usepackage{amsmath}
\usepackage{amssymb}
\usepackage{bm}
\usepackage{amsfonts}
\usepackage[usenames, dvipsnames]{xcolor}
\usepackage{appendix}
\usepackage[colorlinks=true, pdfstartview=FitV, linkcolor=blue, citecolor=blue, urlcolor=blue]{hyperref}
\usepackage{color}
\usepackage{epstopdf}
\usepackage{bbm}
\renewcommand{\Re}{\operatorname{Re}}
\renewcommand{\Im}{\operatorname{Im}}

\begin{document}
 
\title{Dissipation-engineered family of nearly dark states in many-body cavity-atom systems}
 
\author{Rui Lin}
\affiliation{Institute for Theoretical Physics, ETH Z\"urich, 8093 Zurich, Switzerland}
\author{Rodrigo Rosa-Medina}
\affiliation{Institute for Quantum Electronics, ETH Z\"urich, 8093 Zurich, Switzerland}
\author{Francesco Ferri}
\affiliation{Institute for Quantum Electronics, ETH Z\"urich, 8093 Zurich, Switzerland}
\author{Fabian Finger}
\affiliation{Institute for Quantum Electronics, ETH Z\"urich, 8093 Zurich, Switzerland}
\author{Katrin Kroeger}
\affiliation{Institute for Quantum Electronics, ETH Z\"urich, 8093 Zurich, Switzerland}
\author{Tobias Donner}
\affiliation{Institute for Quantum Electronics, ETH Z\"urich, 8093 Zurich, Switzerland}
\author{Tilman Esslinger}
\affiliation{Institute for Quantum Electronics, ETH Z\"urich, 8093 Zurich, Switzerland}
\author{R. Chitra}
\affiliation{Institute for Theoretical Physics, ETH Z\"urich, 8093 Zurich, Switzerland}
\date{\today}

 
\begin{abstract}
 Three-level atomic systems coupled to light have the capacity to host dark states. We study a system of V-shaped three-level atoms coherently coupled to the two quadratures of a dissipative cavity. The interplay between the atomic level structure and dissipation makes the phase diagram of the open system drastically different from the closed one. In particular, it leads to the stabilization of a continuous family of dark and nearly dark excited many-body states with inverted atomic populations as the steady states. The multistability of these states can be probed via their distinct fluctuations and excitation spectra, as well as the system's Liouvillian dynamics which are highly sensitive to ramp protocols. Our model can be implemented experimentally by encoding the two higher-energy modes in orthogonal density-modulated states in a bosonic quantum gas. This implementation offers prospects for potential applications like the realization of quantum optical random walks and microscopy with subwavelength spatial resolution.
\end{abstract}

\maketitle

The interaction between matter and light has received enduring attention over decades. Particularly, dark states can be achieved where atoms are decoupled from the light radiation channel. Single particle dark states lie at the core of  diverse phenomena and applications like coherent population trapping~\cite{arimondo76,radmore82,bergmann98,arimondo96}, electromagnetically induced transparency~\cite{boller91,fleischhauer05}, atomic clocks~\cite{vanier05,arimondo10}, atom cooling~\cite{aspect88,morigi00}, and slow-light polaritons~\cite{fleischhauer02,kupchak15,grusdt16}.
More recently, many-body dark states have been explored, revealing their importance in quantum information and quantum computation~\cite{stannigel12,pichler15,buonaiuto19,pistorius20,cantu20}.

Concurrently, ultracold atomic gases in high-finesse optical cavities have emerged as  a versatile platform to  simulate hitherto unexplored strongly coupled light-matter phases ~\cite{baumann10,ritsch13,mivehvar21}.  A paradigmatic example  is the realization of the Dicke  superradiant phase~\cite{dicke54,hepp73,wang08,carmichael73} in a weakly interacting Bose-Einstein condensate (BEC) coupled to a cavity~\cite{baumann10,klinder15}. 
The ubiquitous dissipation present in these systems can be exploited to obtain squeezing and entanglement~\cite{buca19},  chiral states \cite{dogra19}, as well as oscillatory and chaotic  dynamics~\cite{piazza15,molignini18,iemini18,chiacchio19,lin202,kessler19,zupancic19,kessler20,stitely20,kessler21,prazeres21}. Particularly, cavity dissipation is known to stabilize excited eigenstates as steady states in  the interpolating Dicke{\textendash}Tavis-Cummings (IDTC) model where two-level atoms are coupled to both quadratures of the cavity field~\cite{soriente18,soriente21}, which has recently been experimentally verified by coupling thermal atoms~\cite{zhang18} or a spinor BEC~\cite{ferri21} to an optical cavity. An exciting but relatively underexplored frontier in cavity-QED systems is many-body dark-state physics~\cite{hayn11,emary13,torre13}.

\begin{figure}[!t]
	\centering
	\includegraphics[width=\columnwidth]{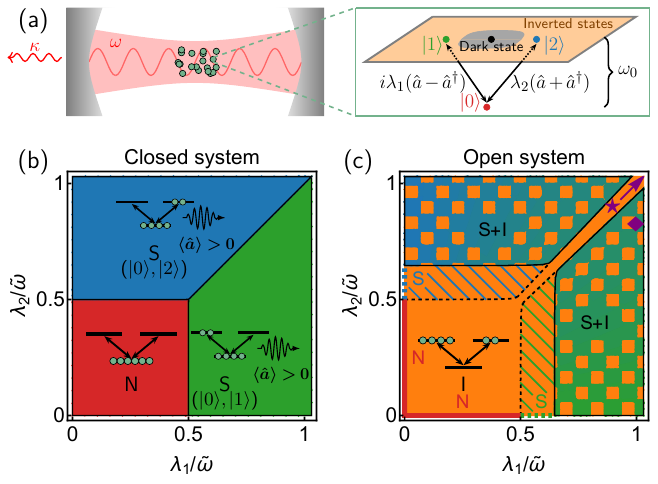}
	\caption{
		(a) System schematics illustrating
		an ensemble of three-level ($|0\rangle$, $|1\rangle$ and $|2\rangle$) atoms with energy splitting $\omega_0$ coupled to an optical cavity $\hat{a}$ with resonance frequency $\omega$ and dissipation rate $\kappa$ through coupling strengths $\lambda_{1,2}$.  Also shown is a representation of the inverted states as mixed/superposition states of the $|1\rangle$ and $|2\rangle$ levels, including the nearly dark (gray region) and the dark (black point) states.
		(b,c) Phase diagrams illustrating (b) the ground states for the closed system Eq.~\eqref{eq:hamil}, and (c) the steady states for the open system Eq.~\eqref{eq:liouvillian}. Pictorial representations of the normal (N), superradiant (S) and inverted (I) states are superimposed.
		In  panel (c), the thick solid and dotted lines indicate that the normal and superradiant states are stable only for $\lambda_1=0$ or $\lambda_2=0$.  Superradiant states stably coexist with the inverted states in the tiled regions, and are physical but unstable in the hatched regions.
		The purple star and diamond indicate the end points of the ramps in Fig.~\ref{fig:inverted_dynamics} and Fig.~\ref{fig:dynamics}, respectively.
		System parameters are $\omega=2\tilde{\omega}$, $\omega_0=0.5\tilde{\omega}$ and $\kappa=0.1\tilde{\omega}$ with reference frequency $\tilde{\omega}$.
	}
	\label{fig:schematics}
\end{figure}

In this work, we study a many-body cavity-atom system where the atomic subspace has an enlarged symmetry, and unveil how dissipative stabilization of excited states fosters the realization of a continuous family of dark and nearly dark steady states with intrinsic many-body correlations. This dark-state preparation is via cavity dissipation in contrast to spontaneous atomic emission~\cite{aspect89,fleischhauer05}.
To this end, we consider $N$ identical, effective V\nobreakdash-shaped three-level atoms coupled to a dissipative cavity with resonance frequency $\omega$ and dissipation rate  $\kappa$ in the thermodynamic limit $N\to\infty$ [Fig.~\ref{fig:schematics}(a)]. 
The atoms have two distinct but degenerate levels $|1\rangle$ and $|2\rangle$ separated by an energy $\omega_0$ from the lowest level $|0\rangle$. The transitions between the ground level and the excited levels are exclusively mediated by coherent couplings to the two orthogonal quadratures of the cavity fields with respective strengths $\lambda_1$ and $\lambda_2$.
In atomic gases, such an effective atomic spectrum can be designed by addressing motional degrees of freedom~\cite{li21,exp,skulte21,kongkhambut21} with external laser fields, or by combining them with internal atomic levels~\cite{fan20}.
The Hamiltonian governing this system is given by ($\hbar=1$)
\begin{eqnarray}\label{eq:hamil}
	\hat{H} &=& \omega \hat{a}^\dagger \hat{a} + \omega_0 (\hat{\Sigma}_{11} + \hat{\Sigma}_{22})+ 
	\frac{i\lambda_1}{\sqrt{N}}(\hat{a} - \hat{a}^\dagger)(\hat{\Sigma}_{01} +\hat{\Sigma}_{10})   \nonumber\\
	&& + \frac{i\lambda_2}{\sqrt{N}}(\hat{a} + \hat{a}^\dagger)(\hat{\Sigma}_{02}-\hat{\Sigma}_{20}),
\end{eqnarray}
where $\hat{a}$ is the cavity annihilation operator, and ${\hat{\Sigma}_{\mu\nu} = \sum_{j=1}^{N}|\mu\rangle_j\langle\nu|_j}$ are the collective pseudospin operators with $|\mu\rangle_j$ denoting the $\mu$th level ($\mu\in\{0,1,2\}$) of the $j$th atom.
A similar model has been considered in Ref.~\cite{fan20}, which focused on the superradiant features in the low energy sectors. 

The Hamiltonian possesses a $\mathbb{Z}_2\times\mathbb{Z}_2$ parity symmetry $\Pi=\mathcal{T}_1\circ\mathcal{T}_2$, where $\mathcal{T}_1$ and $\mathcal{T}_2$ are defined by
$(\hat{a},\hat{\Sigma}_{01},\hat{\Sigma}_{02})\overset{\mathcal{T}_1}{\mapsto}(-\hat{a}^\dagger,\hat{\Sigma}_{01},-\hat{\Sigma}_{02})$ and  $(\hat{a},\hat{\Sigma}_{01},\hat{\Sigma}_{02})\overset{\mathcal{T}_2}{\mapsto}(\hat{a}^\dagger,-\hat{\Sigma}_{01},\hat{\Sigma}_{02})$, and can be broken separately.  When $\lambda_1=\lambda_2$, this symmetry is enlarged to a $\mathrm{U}(1)$ symmetry  with generator $\mathcal{G}=\hat{a}^\dagger \hat{a} - (\hat{\Sigma}_{12}+\hat{\Sigma}_{21})$. The levels $|1\rangle$ and $|2\rangle$ have completely equivalent roles in the Hamiltonian.
The pseudospin operators $\hat{\Sigma}_{\mu\nu}$ follow the commutation relation of the Gell-Mann matrices, and thus span an $\mathrm{SU}(3)$ symmetry space [see Supplementary Material (SM)~\cite{supmat}]. 
In comparison, a spin-1 implementation with $\mathrm{SU}(2)$ symmetry~\cite{zhiqiang17} realizes an equally spaced $\Xi$\nobreakdash-shaped three-level system, where the middle level is equally coupled to the upper and lower ones~\cite{supmat}. This is qualitatively different from our V\nobreakdash-shaped system.

The closed system phase diagram as summarized in Fig.~\ref{fig:schematics}(b) and the corresponding polaritonic excitation spectra can be obtained by using an $\mathrm{SU}(3)$ generalization of the Holstein-Primakoff transformation~\cite{wagner75,supmat}.
For small couplings ${\lambda\equiv\max(\lambda_1,\lambda_2)<\lambda_c=\frac{1}{2}\sqrt{\omega\omega_0}}$, the system stays in the normal phase  with  an empty cavity and all atoms populating the $|0\rangle$ level. When either coupling exceeds the threshold $\lambda>\lambda_c$, the system enters the superradiant phase where the cavity field is coherently populated as ${|\langle \hat{a}\rangle|=\frac{\lambda\sqrt{N}}{\omega}\sqrt{1-\left(\frac{\omega_0\omega}{4\lambda^2}\right)^2}}$.  
For ${\lambda_1>\lambda_2}$ (${\lambda_2>\lambda_1}$), the $\mathcal{T}_1$ ($\mathcal{T}_2$) symmetry is spontaneously broken, leading to a nonzero expectation value of the  imaginary (real) quadrature of the cavity field and an occupation of  the $|0\rangle$ and $|1\rangle$ ($|2\rangle$) levels.
For ${\lambda_1=\lambda_2 > \lambda_c}$, the broken $\mathrm{U}(1)$ symmetry results in a population of all three atomic levels.

In the high-energy sector, our model hosts a dark state decoupled from the cavity field and thus stable, obeying $\hat{H}|D\rangle=N\omega_0|D\rangle$ and  ${\hat{a}|D\rangle=0}$~\cite{kraus08,diehl08,finkelstein19}:
\begin{eqnarray}\label{eq:dark_state}
	|D \rangle= \prod_{j=1}^{N}\frac{\lambda_2|1\rangle_j+ \lambda_1 |2\rangle_j}{\sqrt{\lambda_1^2+\lambda_2^2}}.
\end{eqnarray}
This state manifests a complete atomic population inversion with unoccupied $|0\rangle$ level. In fact, it is merely one element of a family of states satisfying $\langle \hat{\Sigma}_{0\nu}\rangle=0$ and $\langle \hat{a}\rangle=0$ in $\mathcal{O}(N)$, which we term the \emph{inverted states}. These states are uniquely defined by two parameters,
\begin{eqnarray}\label{eq:def_inverted}
	N_1 = \langle \hat{\Sigma}_{11}\rangle,\quad \theta = \arg\langle\hat{\Sigma}_{12}\rangle,
\end{eqnarray}
where $N_1\in[0,N]$ is the occupation of the $|1\rangle$ level, and $\theta\in(-\pi,\pi]$ is the relative phase between the $|1\rangle$ and $|2\rangle$ levels. These quasidegenerate inverted states have a much higher energy ${E=N\omega_0}$ compared to the polaritonic excitations, whose energy is of $\mathcal{O}(1)$~\cite{supmat}. These states stem from the enlarged $\mathrm{SU}(3)$ symmetry and, in contrast to the dark state of Eq.~\eqref{eq:dark_state}, are characterized by nontrivial many-body correlations.
Despite the inaccessibility of these inverted states with quasiadiabatic protocols in the closed system, they manifest a nontrivial relation to the dark state upon introduction of dissipation.

We now explicitly consider cavity dissipation via a Liouvillian time evolution of the density matrix~\cite{dimer07},
\begin{eqnarray}\label{eq:liouvillian}
	\partial_t \hat{\rho} = \mathcal{L}\hat{\rho} = -i[\hat{H},\hat{\rho}] + \kappa (2\hat{a}\hat{\rho} \hat{a}^\dagger - \{\hat{a}^\dagger \hat{a},\hat{\rho}\}).
\end{eqnarray}
The stability of the Liouvillian's fixed points determines the steady states of the system. In the third quantization approach~\cite{prosen08,prosen10}, this can be inferred from \emph{rapidities} $\{\xi_i\}$, whose construction and calculation from the Liouvillian are detailed in SM~\cite{supmat}. The values of $\xi_i$ dictate the stability of the steady states to fluctuations. A state is stable (unstable) when ${\Re\xi_\mathrm{min}\ge0}$ (${\Re\xi_\mathrm{min}< 0}$), with $\xi_\mathrm{min}$ the rapidity with the minimal real part. 
The system converges to (deviates from) it with a rate of $2|\Re\xi_\mathrm{min}|$ and an oscillation frequency of $2|\Im\xi_\mathrm{min}|$~\cite{prosen08,prosen10}. As detailed in SM~\cite{supmat}, although the normal, superradiant, and inverted states all remain fixed points of the Liouvillian [Eq.~\eqref{eq:liouvillian}], their stabilities are dramatically altered by dissipation. The resulting open system stability diagram is summarized in Fig.~\ref{fig:schematics}(c).  Without loss of generality, we assume $\lambda_1\ge\lambda_2$ in the following discussion.

The first intriguing  aspect of the open system is the generic instability of the normal state, which is now stable only when $\lambda_2=0$ recovering the Dicke model, and $\lambda_1<\frac{1}{2}\sqrt{(\omega^2+\kappa^2)\omega_0/\omega}$~\cite{fan20}.
Dissipation also significantly impacts the superradiant states, where all atomic levels are now populated. At the level of fixed points, akin to the IDTC model~\cite{soriente18}, the $\mathrm{U}(1)$ symmetry broken phase at $\lambda_1=\lambda_2$ is eliminated and two superradiant boundaries separated by a sliver emerge symmetric about $\lambda_1=\lambda_2$. Each superradiant boundary harbors both continuous and first order sections~\cite{supmat}. In contrast to the IDTC model, the superradiant state is unstable in an intermediate region above the critical coupling of the closed system [hatched region in Fig.~\ref{fig:schematics}(c)], as inferred from the associated rapidities~\cite{supmat}.  This superradiant state stability boundary is almost insensitive to the weaker coupling,  and lies where the stronger coupling $\lambda_1$ takes the approximate value of
\begin{eqnarray}\label{eq:sr_stable_boundary_open}
	\lambda_\mathrm{stable} \approx \sqrt{(\omega^2+\kappa^2)(\omega+2\sqrt{4\omega^2+3\kappa^2})\omega_0/12\omega^2}.
\end{eqnarray}
Both normal and superradiant states manifest mathematical singularity in fluctuations in both limits ${\lambda_2\to0}$ and ${\kappa\to0}$~\cite{supmat}. A slight deviation from the Dicke model together with an infinitesimal dissipation immediately destabilizes the superradiant state when ${1\le\lambda_1/\lambda_c\lesssim\sqrt{5/3}}$ and the normal state.

\begin{figure}[t]
	\centering
	\includegraphics[width=\columnwidth]{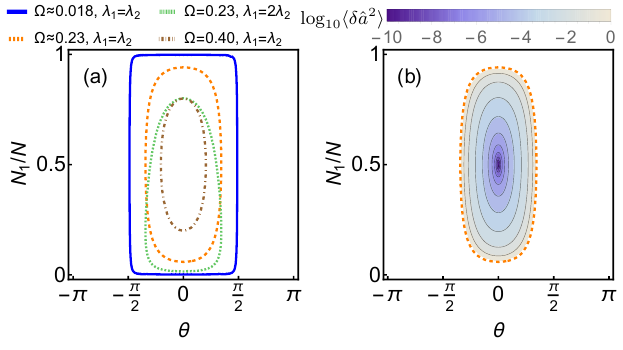}
	\caption{
		(a) Stability boundaries of the inverted state in the open system for different system parameters given by  Eq.~\eqref{eq:inverted_stable_boundary_open}. The  regions enclosed indicate stable solutions. The boundary is only sensitive to the coupling ratio $\lambda_1/\lambda_2$ and the scale variable $\Omega$ . The blue solid line corresponds to typical experimental parameters $\omega=2\pi\times2.0$~MHz, $\omega_0=2\pi\times50$~kHz and $\kappa=2\pi\times1.25$~MHz~\cite{ferri21,exp}, the orange dashed line and green dotted line correspond to the parameters used in Fig.~\ref{fig:schematics}: $\omega=2\tilde{\omega}$, $\omega_0=0.5\tilde{\omega}$, $\kappa=0.1\tilde{\omega}$, whereas the brown dash-dotted line depicts a scenario with larger $\Omega=0.4$. 
		(b) Cavity field fluctuations $\langle \delta\hat{a}^2\rangle$ [cf. Eq.~\eqref{eq:cav_fluc}] of different inverted states in the stable region, with system parameters taken as $\omega=2\tilde{\omega}$, $\omega_0=0.5\tilde{\omega}$, $\kappa=0.1\tilde{\omega}$, and $\lambda_1=\lambda_2=\tilde{\omega}$.
	}
	\label{fig:inverted}
\end{figure}

The high-energy inverted states show a nontrivial stability, as only a subset of them is stable.
In the $N_1$\nobreakdash-$\theta$ parameter space, the inverted state stability boundary as depicted in Fig.~\ref{fig:inverted}(a) is given by
\begin{eqnarray}\label{eq:inverted_stable_boundary_open}
	\frac{\eta_1\eta_2 \cos\theta}{\eta_1^2+\eta_2^2} = \frac{\omega\omega_0}{\omega^2+\omega_0^2+\kappa^2}\equiv\Omega,
\end{eqnarray}
where ${\eta_1 = \lambda_1 \sqrt{N_1/N}}$, ${\eta_2=\lambda_2\sqrt{1-N_1/N}}$, and ${\Omega\in(0,1/2)}$ is a scaled variable. 
The enclosed extended region of multistability has a finite area ${A=N\pi\left[1-\Omega(\lambda_1^2+\lambda_2^2)/\sqrt{\lambda_1^2\lambda_2^2+(\lambda_1^2-\lambda_2^2)^2\Omega^2}\right]}$,
which increases for less resonant $\omega$ and $\omega_0$, and for larger $\kappa$.
Consistently, the dark state $|D\rangle$ introduced in Eq.~\eqref{eq:dark_state} corresponds to ${(N_1,\theta)=\left(\frac{N\lambda_2^2}{\lambda_1^2+\lambda_2^2},0\right)}$, and always lies inside the stable region for all values of $\Omega$. These \emph{nearly dark} inverted states are either the
exclusive steady states [orange region in Fig.~\ref{fig:schematics}(c)] or coexistents with superradiant states [hatched regions in Fig.~\ref{fig:schematics}(c)]. 
We reiterate that such multistable steady states cannot be realized using $\mathrm{SU}(2)$ atoms with cavity modes coupled linearly to it (see SM~\cite{supmat}), which can host only one coherent dark state~\cite{emary13,torre13}. Indeed, their realization requires a larger atomic symmetry like $\mathrm{SU}(3)$~\cite{supmat}, and has been predicted in an $\mathrm{SU}(3)$ atomic system coupled to two cavity modes~\cite{hayn11}.
Moreover, their existence and population inversion further requires both the degeneracy of the $|1\rangle$ and $|2\rangle$ levels and a positive $\omega_0$ as well.  For $\omega_0<0$, a similar family of nearly dark states exists albeit without population inversion (see SM~\cite{supmat}). An indepth exploration beyond these parameter regimes merits future study.

Our results motivate further questions: (i) Are there accessible observables physically distinguishing different stable inverted states? (ii) Which of these states does the system converge to during its Liouvillian time evolution?  

To distinguish between the inverted states, a direct measurement of the atomic observables $N_1$ and $\theta$ can be experimentally challenging.  An alternative is to extract the cavity and atomic fluctuations as well as the excitation spectra harbored by the individual states. Particularly, the cavity fluctuations $\langle \delta \hat{a}^2\rangle \equiv \langle \hat{a}^\dagger \hat{a}\rangle - |\langle \hat{a} \rangle|^2$ are found to be~\cite{supmat}
\begin{widetext}
\begin{eqnarray}\label{eq:cav_fluc}
	\langle \delta \hat{a}^2\rangle = \frac{(\eta_1^4+\eta_2^4-2\eta_1^2\eta_2^2\cos2\theta)(\omega^2+\kappa^2+4\eta_1\eta_2\cos\theta)\omega_0^2}{2(8\eta_1^2\eta_2^2(1+\cos2\theta) + 4(\eta_1^2+\eta_2^2)\omega\omega_0 + (\omega^2+\kappa^2)\omega_0^2)(\eta_1\eta_2(\omega^2+\omega_0^2+\kappa^2)\cos\theta-(\eta_1^2+\eta_2^2)\omega\omega_0)},
\end{eqnarray}
and vary over orders of magnitude within the stable region, diverging at the stability boundary and strongly suppressed around $|D\rangle$ [Fig.~\ref{fig:inverted}(b)]. 
The vanishing cavity and atomic fluctuations at $|D\rangle$ corroborate its darkness and atomic coherence, whereas the states in its vicinity are mixed states with intrinsic many-body correlations and finite fluctuations.
Measurements of the cavity fluctuations and the excitation spectra can uniquely determine the inverted state for unequal couplings ${\lambda_1\neq\lambda_2}$, but only up to a closed contour in the $N_1$\nobreakdash-$\theta$ parameter space for equal couplings ${\lambda_1=\lambda_2}$, as consistent with the  $\mathrm{U}(1)$ symmetry~\cite{supmat}.
\pagebreak
\end{widetext}

\begin{figure}[t]
	\centering
	\includegraphics[width=\columnwidth]{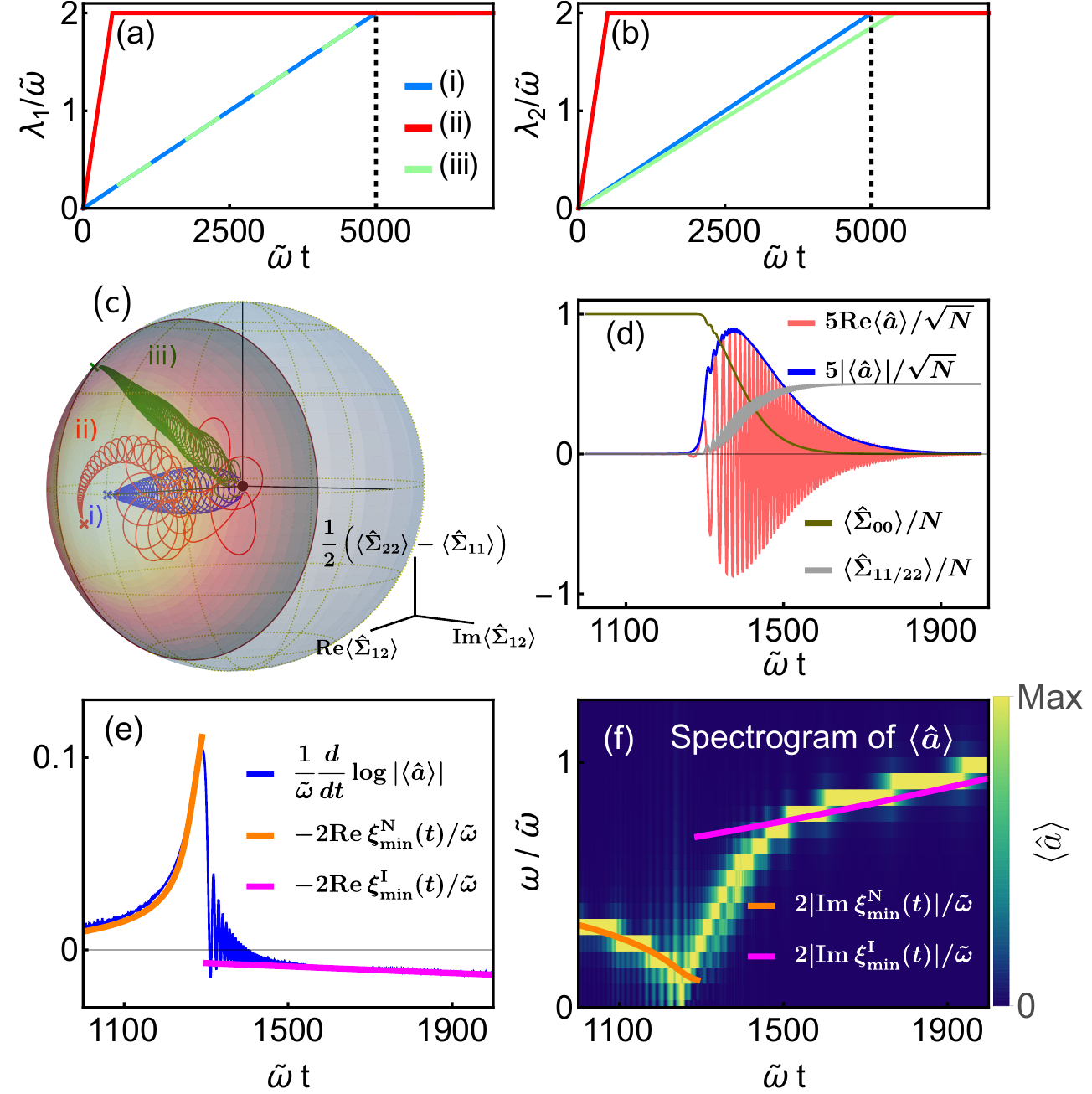}
	\caption{  
		Dynamical evolution probing the multistability of the inverted states. (a,b) The time-dependence of (a) $\lambda_1(t)$ and (b) $\lambda_2(t)$ in the three protocols~\cite{supmat}.
		(c) The trajectories of all three protocols projected on the Bloch sphere spanned by $|1\rangle$ and $|2\rangle$ levels. The black dot indicates the starting point of the trajectories, i.e., the normal state, while the crosses indicate the final points. The stable region of the inverted states, appearing as the colored spherical cap in this representation~\cite{supmat}, corresponds to the one shown in Fig.~\ref{fig:inverted}(b).
		(d) Evolution of the cavity and atomic fields during the pulse in protocol (i), where (e,f) the simulated dynamics is consistent with the theoretically calculated rapidities $\xi_\mathrm{min}^\mathrm{N}$ for normal state and $\xi_\mathrm{min}^\mathrm{I}$ for inverted state with $(N_1,\theta)=(1/2,0)$ evaluated with instantaneous coupling strengths. Particularly, (e) the time derivative of $|\langle\hat{a}\rangle|$ depicting the deviation/convergence rate  agree quantitatively to $\Re\xi_\mathrm{min}$, whereas (f) the spectrogram of $\langle\hat{a}\rangle$ depicting the oscillation frequency to $\Im\xi_\mathrm{min}$.
	}
	\label{fig:inverted_dynamics}
\end{figure}

The full Liouvillian dynamics of the system can be captured by numerically solving the coupled mean-field equations of motion for the cavity and atomic fields, using the normal state with a small cavity field as the initial state, and different time-dependent ramp protocols for the two couplings. These are seven coupled complex equations governing  the expectation values of  $\langle  \hat{a}\rangle$, $\langle \hat{\Sigma}_{01}\rangle$, $\langle\hat{\Sigma}_{02}\rangle$, $\langle \hat{\Sigma}_{12}\rangle$, $\langle \hat{\Sigma}_{00}\rangle$, $\langle \hat{\Sigma}_{11}\rangle$ and $\langle \hat{\Sigma}_{22}\rangle$~\cite{supmat}. As a representative case, the system parameters are chosen as $\omega=2\tilde{\omega}$, $\omega_0=0.5\tilde{\omega}$ and $\kappa=0.1\tilde{\omega}$ with reference frequency $\tilde{\omega}$.

We ramp up the couplings from ${\lambda_1=\lambda_2=0}$ to ${\lambda_1 = \lambda_2=2\tilde{\omega}}$ using three different protocols as illustrated in Fig.~\ref{fig:inverted_dynamics}(a,b), which differ in ramp rate and path in $\lambda_1$\nobreakdash-$\lambda_2$ parameter space. 
For a better visualization, the ensuing Liouvillian trajectories are projected onto the Bloch sphere spanned by the axes $\Re\langle \hat{\Sigma}_{12}\rangle$, $\Im\langle \hat{\Sigma}_{12}\rangle$ and $\frac{1}{2}\left(\langle \hat{\Sigma}_{22}\rangle-\langle \hat{\Sigma}_{11}\rangle\right)$ [Fig.~\ref{fig:inverted_dynamics}(c)]. Despite identical final couplings, the final converged state depends sensitively on both ramp rate and path, signaling the multistability of the inverted states.
The nature of the dynamics is further elucidated by studying the cavity field evolution. As the atomic population inverts, correlations between atomic levels $\langle\hat{\Sigma}_{01}\rangle$ and $\langle\hat{\Sigma}_{02}\rangle$ are established. This automatically generates a nonzero $\langle \hat{a}\rangle$ signifying a burst of photons [Fig.~\ref{fig:inverted_dynamics}(d)]. We can best understand this in the bad-cavity limit ${\kappa\gg\omega_0}$, where the cavity field follows the atomic evolution adiabatically as
${\langle \hat{a}\rangle = \left(i\lambda_1\Re\langle\hat{\Sigma}_{01}\rangle +  \lambda_2\Im\langle\hat{\Sigma}_{02}\rangle\right)/[\sqrt{N}(\omega+i\kappa)]}$. 
The quantitative consistency between cavity field dynamics and rapidities [Fig.~\ref{fig:inverted_dynamics}(e,f)] confirms the dissipative nature of the instability driving the population inversion.

\begin{figure}[t]
	\centering
	\includegraphics[width=\columnwidth]{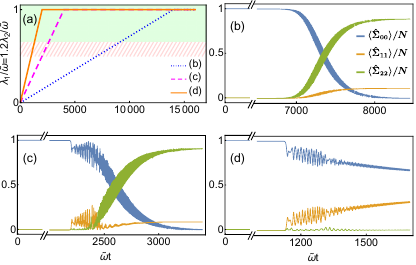}
	\caption{
		Dynamical evolution probing the accessibility of the superradiant states. (a) The ramp protocols of the coupling strengths~\cite{supmat}. The regions in which the superradiant state is unstable, stable, and unphysical for the corresponding instantaneous coupling strengths are marked in green, hatched red, and white, respectively, cf. Fig.~\ref{fig:schematics}(c). (b-d)  Qualitatively different evolutions of the atomic fields for  the three protocols. 
	}
	\label{fig:dynamics}
\end{figure}

We now discuss the accessibility of the superradiant steady states by considering three ramp protocols satisfying $\lambda_1/\lambda_2=1.2$ [Fig.~\ref{fig:dynamics}(a)]. They traverse the unstable superradiant region, and terminate in the region where both the superradiant state and the inverted states are stable [purple diamond in Fig.~\ref{fig:schematics}(c)]. We find a ramp-dependent dynamics  mirroring the complex stability of the states, as shown in Fig.~\ref{fig:dynamics}.  For the slowest ramp [Fig.~\ref{fig:dynamics}(b)], the dynamics is dominated by the instability of the normal state to the inverted state. 
For an intermediate ramp rate [Fig.~\ref{fig:dynamics}(c)], the  system first enters the  unstable superradiant state  before being driven by its instability to the inverted state.
Finally, for a fast enough ramp [Fig.~\ref{fig:dynamics}(d)], the system is quenched to the stable superradiant state before it can invert towards the nearly dark states.

\pagebreak

Our model can be experimentally implemented using a two-dimensional BEC in the $x$\nobreakdash-$z$ plane with effectively two internal Zeeman sublevels ${|m=0\rangle}$ and ${|m=1\rangle}$  coupled to a dissipative cavity with typical parameters of ${\omega=2\pi\times2.0}$~MHz and ${\kappa=2\pi\times1.25}$~MHz, and driven by a bichromatic laser whose two standing-wave modulations are phase-shifted by $\pi/2$ at the position of the atomic cloud~\cite{exp}. To the lowest order in kinetic energy, this atomic system can be effectively mapped to our model [Eq.~\eqref{eq:hamil}], where $|0\rangle$ corresponds to a spatially uniform state ${\psi_0\propto |m=0\rangle\otimes1}$, while $|1\rangle$ and $|2\rangle$ are orthogonal spatially modulated modes with wavevector $k$: ${\psi_1 \propto |m=1\rangle\otimes\cos(kx)\cos(kz)}$ and ${\psi_2 \propto |m=1\rangle\otimes\cos(kx)\sin(kz)}$. This implementation structurally protects the degeneracy of the $|1\rangle$ and $|2\rangle$ levels. The energy difference $\omega_0$ between the atomic levels is contributed by both the Zeeman splitting and the recoil energy, and has a typical value of ${\omega_0=2\pi\times50}$~kHz. Controlled by the pump laser, the couplings $\lambda_1$ and $\lambda_2$ take values in the range of $2\pi\times100$~kHz. For these experimentally associated parameters, the inverted state stability boundary in $N_1$\nobreakdash-$\theta$ parameter space is plotted as the blue solid curve in Fig.~\ref{fig:inverted}(a), showing a vast multistable region and thus indicating an easy observability of our predicted results. Other proposed experimental realizations of similar models are also discussed in Refs.~\cite{fan20,kongkhambut21}.

In conclusion, the dissipative stabilization of a continuous family of excited many-body states as steady states establishes a new paradigm for preparing nearly dark states in cavity-atom systems. These salient features pave the way for  a wide range of prospective applications. For instance, the large multistable region provides a potential platform for implementing fluctuation-driven random walks like L\'{e}vy flight~\cite{levy54}, which can be used for atom cooling~\cite{bertin08,bardou01,rocha20}.
Moreover, in the experimental implementation discussed above~\cite{exp}, the established correspondence between matter and light  can potentially be used  for fluctuation-based microscopy of the atomic density patterns with subwavelength resolution~\cite{hawkes10}.

\begin{acknowledgments}
	We are grateful to M. Landini, N. Dogra, A.U.J. Lode and O. Zilberberg for fruitful discussions.  R.L and R.C acknowledge funding from the ETH grants. R.R-M, F. Ferri, F. Finger, T.D, and T.E acknowledge funding from the Swiss National Science Foundation: project numbers 182650 and 175329 (NAQUAS QuantERA) and NCCR QSIT, from EU Horizon2020: ERC Advanced grant TransQ (project Number 742579).
\end{acknowledgments}

\bibliographystyle{apsrev}
\bibliography{References}

\end{document}


\title{Supplementary Material: \\ Dissipation-engineered family of nearly dark many-body states in cavity-atom systems}

\author{Rui Lin}
\affiliation{Institute for Theoretical Physics, ETH Z\"urich, 8093 Zurich, Switzerland}
\author{Rodrigo Rosa-Medina}
\affiliation{Institute for Quantum Electronics, ETH Z\"urich, 8093 Zurich, Switzerland}
\author{Francesco Ferri}
\affiliation{Institute for Quantum Electronics, ETH Z\"urich, 8093 Zurich, Switzerland}
\author{Fabian Finger}
\affiliation{Institute for Quantum Electronics, ETH Z\"urich, 8093 Zurich, Switzerland}
\author{Katrin Kroeger}
\affiliation{Institute for Quantum Electronics, ETH Z\"urich, 8093 Zurich, Switzerland}
\author{Tobias Donner}
\affiliation{Institute for Quantum Electronics, ETH Z\"urich, 8093 Zurich, Switzerland}
\author{Tilman Esslinger}
\affiliation{Institute for Quantum Electronics, ETH Z\"urich, 8093 Zurich, Switzerland}
\author{R. Chitra}
\affiliation{Institute for Theoretical Physics, ETH Z\"urich, 8093 Zurich, Switzerland}

 
\maketitle
{\hypersetup{linkcolor=black}
\tableofcontents
}
\newpage
\section{Symmetry of the atomic system}

\subsection{Mapping to the $\mathrm{SU}(3)$ Gell-Mann matrices}

We now briefly discuss the underlying symmetry of the atomic operators. In the Hamiltonian of our model [Eq.~(1) of the main text]~\footnote{Throughout this Supplementary Material, we omit the hat symbol on operators $\hat{O}\to O$ for clarity of notations.}
\begin{eqnarray}\label{eq:hamil_suppmat}
	H = \omega a^\dagger a + \omega_0 \Sigma_{11} + \omega_0\Sigma_{22}
	+ \frac{i\lambda_1}{\sqrt{N}}(a - a^\dagger)(\Sigma_{01}+\Sigma_{10}) 
	+ \frac{i\lambda_2}{\sqrt{N}}(a + a^\dagger)(\Sigma_{02}-\Sigma_{20}),
\end{eqnarray}
the atoms are in an $\mathrm{SU}(3)$ subspace. This can be revealed by the relation between the collective atomic operators $\Sigma_{\mu\nu}$ to the Gell-Mann matrices $g_i$ with $i=1,2,\dots,8$, which are the generators of the $\mathrm{SU}(3)$ symmetry spanning the $\mathfrak{su}(3)$ Lie algebra. Together with the identity matrix $g_0=\mathbbm{1}_{3\times3}$, these Gell-Mann matrices,
\begin{align}
	\begin{split}
	&g_1 = \begin{pmatrix}
		0&1&0\\1&0&0\\0&0&0
	\end{pmatrix},\quad
	g_2 = \begin{pmatrix}
		0&-i&0\\i&0&0\\0&0&0
	\end{pmatrix},\quad
	g_3 = \begin{pmatrix}
		1&0&0\\0&-1&0\\0&0&0
	\end{pmatrix},\quad
	g_4 = \begin{pmatrix}
		0&0&1\\0&0&0\\1&0&0
	\end{pmatrix},\\
	&g_5 = \begin{pmatrix}
		0&0&-i\\0&0&0\\i&0&0
	\end{pmatrix},\quad
	g_6 = \begin{pmatrix}
		0&0&0\\0&0&1\\0&1&0
	\end{pmatrix},\quad
	g_7 = \begin{pmatrix}
		0&0&0\\0&0&-i\\0&i&0
	\end{pmatrix},\quad
	g_8 = \frac{1}{\sqrt{3}}\begin{pmatrix}
		1&0&0\\0&1&0\\0&0&-2
	\end{pmatrix},
	\end{split}
\end{align}
form a complete basis for $3\times3$ matrices. Notably, within the $\mathfrak{su}(3)$ algebra, there are an infinite number of sets of Gell-Mann matrices constituting $\mathfrak{su}(2)$ subalgebras, examples being $\{g_1,g_2,g_3\}$, $\{g_4,g_5,g_3/2+\sqrt{3}g_8/2\}$, $\{g_6,g_7,-g_3/2+\sqrt{3}g_8/2\}$, etc. If the Hamiltonian only contains terms which are linear combinations of matrices from a chosen $\mathfrak{su}(2)$ subalgebra, the system symmetry is reduced to $\mathrm{SU}(2)$.

With our Hamiltonian Eq.~\eqref{eq:hamil_suppmat}, we can find a mapping from the atomic operators to the Gell-Mann matrices preserving the commutation relation between the operators:
\begin{align}
	\begin{split}
		\Sigma_{01} + \Sigma_{10} &\mapsto g_1, \\
		i(\Sigma_{02} -\Sigma_{20}) &\mapsto g_5,\\
		\Sigma_{11} &\mapsto -\frac{1}{2}g_3 + \frac{1}{2\sqrt{3}}g_8+\frac{1}{3}g_0, \\
		\Sigma_{22} &\mapsto -\frac{1}{\sqrt{3}}g_8+\frac{1}{3}g_0.
	\end{split}
\end{align}
It is impossible to write these four operators as linear combinations of Gell-Mann matrices from an $\mathfrak{su}(2)$ subalgebra, and we thus conclude that the atomic subspace has an underlying $\mathrm{SU}(3)$ symmetry. 
Nevertheless, in the Dicke limit where $\lambda_2=0$ and thus the $g_5$ term vanishes, the operators can be spanned under the basis ${\{g_0,g_1,g_2,g_3,g_8\}}$. Among these basis matrices, $\{g_1,g_2,g_3\}$ constitute an $\mathfrak{su}(2)$ subalgebra, and all three matrices commute with $g_8$, which by itself can be treated as the generator of a $\mathrm{U}(1)$ symmetry. As a result, the underlying symmetry of our system is reduced to $\mathrm{SU}(2)\times\mathrm{U}(1)$. The same argument also applies for the case $\lambda_1=0$.

\subsection{Conserved quantities in the atomic systems}

An $\mathrm{SU}(3)$ system has three intrinsic conserved quantities guaranteed by the symmetry. The first quantity is the expectation value of $g_0$, whose conservation is guaranteed by the tracelessness of the Gell-Mann matrices $g_i$, $i=1,2,\dots,8$; whereas the other two quantities are the expectation values of the quadratic and cubic Casimirs $C_1$ and $C_2$ of the $\mathrm{SU}(3)$ symmetry, which are defined by
\begin{eqnarray}
	C_1 = \sum_{i=1}^{8} g_ig_i,\quad C_2 = \sum_{i,j,k=1}^{8} d_{i,j,k}g_ig_jg_k,
\end{eqnarray}
respectively, where $d_{i,j,k} = \Tr(g_ig_jg_k+g_jg_ig_k)$. In the atomic system, the condition of tracelessness implies the conservation of total particles in the system
\begin{eqnarray}\label{eq:total_number}
	N = \sum_{\mu=0}^{2}\langle \Sigma_{\mu\mu}\rangle
\end{eqnarray}
whereas the other two conserved quantities are respectively
\begin{subequations}\label{eq:Casimir}
\begin{eqnarray}
	C_1 &=& \sum_{\mu=0}^{2}\langle\Sigma_{\mu\mu}\rangle^2 + \sum_{\{\mu,\nu\}} \Big(3\vert\langle\Sigma_{\mu\nu}\rangle\vert^2-\langle\Sigma_{\mu\mu}\rangle\langle\Sigma_{\nu\nu}\rangle\Big) = N^2 \\
	C_2 &=& \frac{9}{2}\sum_{\{\mu,\nu,\rho\}}\vert\langle\Sigma_{\mu\nu}\rangle\vert^2\Big(\langle \Sigma_{\mu\mu}\rangle+\langle\Sigma_{\nu\nu}\rangle-2\langle\Sigma_{\rho\rho}\rangle\Big)  
	 - \frac{1}{2}\prod_{\{\mu,\nu,\rho\}} \Big(\langle \Sigma_{\mu\mu}\rangle+\langle\Sigma_{\nu\nu}\rangle-2\langle\Sigma_{\rho\rho}\rangle\Big)
	 + 27 \Big\vert \langle \Sigma_{01}\rangle \langle\Sigma_{12}\rangle \langle \Sigma_{20}\rangle \Big\vert
	= N^3. \nonumber\\
\end{eqnarray}
\end{subequations}
where the summation $\sum_{\{\mu,\nu\}}$ runs over the pairs $\{\mu,\nu\}=\{0,1\},\{1,2\},\{2,0\}$, while the summation $\sum_{\{\mu,\nu,\rho\}}$ and the product $\prod_{\{\mu,\nu,\rho\}}$ run over the triplets $\{\mu,\nu,\rho\}=\{0,1,2\},\{1,2,0\},\{2,0,1\}$.

\subsection{Three-level $\mathrm{SU}(2)$ spin-1 systems}

In the main text we have claimed that our system is essentially different from a system with an $\mathrm{SU}(2)$ three-level atomic structure, i.e. a spin-1 atomic structure. Here we substantiate the claim by discussing the general form of the Hamiltonian of such a system, which reads~\cite{zhiqiang17}:
\begin{eqnarray}\label{eq:hamil_spin_1}
	H = \omega a^\dagger a + \omega_0 (\Sigma_{11} - \Sigma_{22})+
	\frac{1}{\sqrt{N}}(\lambda_1a+\lambda_1^*a^\dagger)(\Sigma_{01} +\Sigma_{10}+\Sigma_{02} +\Sigma_{20}) +
	\frac{1}{\sqrt{N}}(\lambda_2a-\lambda_2^*a^\dagger)(\Sigma_{01} - \Sigma_{10}+\Sigma_{02} -\Sigma_{20}).
\end{eqnarray}
This Hamiltonian describes a $\Xi$-shaped system with $|0\rangle$ level in the middle, and $|1\rangle$ and $|2\rangle$ levels separated from it with the same energy difference. Moreover, the couplings from the $|0\rangle$ level to the $|1\rangle$ and $|2\rangle$ levels always share the same strengths.
The operators in the Hamiltonian can be mapped to the Gell-Mann matrices as 
\begin{eqnarray}
	\begin{split}
	\Sigma_{11}-\Sigma_{22} &\mapsto g_3/2+\sqrt{3}g_8/2 \\
	\Sigma_{01} +\Sigma_{10}+\Sigma_{02} +\Sigma_{20} &\mapsto g_1+g_6 \\
	-i(\Sigma_{01} - \Sigma_{10}+\Sigma_{02} -\Sigma_{20}) &\mapsto g_2+g_7
	\end{split}
\end{eqnarray}
These three operators obey the commutation relation of the $\mathfrak{su}(2)$ subalgebra, and thus span the basis for it. We can further identify these operators as the $\mathrm{SU}(2)$ generators $J_z$, $J_x$, and $J_y$, respectively, which in a two-level system can be represented by the $2\times2$ Pauli matrices.

\section{Eigenstates and their stability in the closed system}\label{sec:closed}

In this section, we investigate the eigenstates and their stability in the closed system based on the Holstein-Primakoff (HP) transformation~\cite{wagner75}.
The HP transformation is a useful tool for investigating the fluctuations around a ground state in the closed system. It has two advantages: (i) it focuses on the behaviors of the fluctuations by dropping the averaged mean-field behaviors of the state; and (ii) it yields a quadratic Hamiltonian which can be further solved by the Hopfield-Bogoliubov transformation technique. Combining these two techniques, we can solve the excitation spectrum, find the stability, and calculate the cavity and atomic fluctuations around a given state. 

\subsection{Holstein-Primakoff transformation in the closed system}

We now derive the HP Hamiltonian in the closed system. For a three-level system, the recipe for a HP transformation is given by Ref.~\cite{wagner75}. The two-level counterpart of this technique has been applied on the Dicke and IDTC models in Refs.~\cite{emary03} and \cite{baksic14}, respectively. We first show a detailed derivation for the normal and superradiant states, and then briefly discuss and show the results for the inverted state.

\subsubsection{Normal and superradiant states}

In the normal and superradiant state, the HP transformation is formally given by~\cite{wagner75}
\begin{eqnarray}\label{eq:HP_mapping1}
	\begin{split}
	\Sigma_{\mu0} &\to b_\mu^\dagger\sqrt{N-b_1^\dagger b_1-b_2^\dagger b_2}, \\
	\Sigma_{\mu\mu}&\to b_\mu^\dagger b_\mu,
	\end{split}
\end{eqnarray}
with $\mu=1,2$. The operators $a$ and $b_{1,2}$ are further split into their respective expectation values $\alpha$, $\beta_{1,2}$ and fluctuation operators $c$, $d_{1,2}$:
\begin{eqnarray}
	\begin{split}
	a &= c + \alpha, \\
	b_{1,2} &= d_{1,2} + \beta_{1,2}.
	\end{split}
\end{eqnarray}
The expectations values $\alpha$ and $\beta_{1,2}$ scale as $\mathcal{O}(\sqrt{N})$, while the fluctuations $c$ and $d_{1,2}$ scale as $\mathcal{O}(1)$ in terms of $N$.

After the mapping, the next step is to perform a Taylor expansion on the square root $\sqrt{N-b_1^\dagger b_1-b_2^\dagger b_2}$, while keeping only the terms which scale higher than $\mathcal{O}(1)$ in terms of $N$. Upon this step, we can find the collective total energy of the system which scales in the order of $\mathcal{O}(N)$
\begin{subequations}
	\begin{eqnarray}
		E = \omega |\alpha|^2 + \omega_0 (|\beta_1|^2+|\beta_2|^2) + i  \sqrt{1-\frac{|\beta_1|^2+|\beta_2|^2}{N}}  [\lambda_1(\alpha - \alpha^*)(\beta_1 + \beta_1^*)+
		\lambda_2(\alpha + \alpha^*)(\beta_2 - \beta_2^*)].
	\end{eqnarray}
The mean-field solutions of $\alpha$, $\beta_1$, and $\beta_2$ are the extrema of the total mean-field energy with respect to $\alpha^*$, $\beta_1^*$, and $\beta_2^*$. This leads to three equations
\begin{eqnarray}
	\begin{split}
	&\omega \alpha - i \frac{\lambda_1}{\sqrt{N}} \sqrt{k}  (\beta_1+\beta_1^*) + i \frac{\lambda_2}{\sqrt{N}} \sqrt{k} (\beta_2 - \beta_2^*)=0,\\
	&\omega_0 \beta_1 +i \frac{\lambda_1}{\sqrt{N}} \sqrt{k} (\alpha - \alpha^*) \left[ 1- \frac{\beta_1}{2k} (\beta_1 + \beta_1^*)\right] -i \frac{\lambda_2}{\sqrt{N}} \sqrt{k}(\alpha+\alpha^*)\frac{\beta_1(\beta_2 - \beta_2^*)}{2k}=0, \\
	&\omega_0 \beta_2 +i \frac{\lambda_2}{\sqrt{N}} \sqrt{k}(\alpha + \alpha^*) \left[ -1 - \frac{\beta_2}{2k} (\beta_2 - \beta_2^*)\right] -i \frac{\lambda_1}{\sqrt{N}} \sqrt{k} (\alpha - \alpha^*)\frac{\beta_2(\beta_1 + \beta_1^*)}{2k}=0,
	\end{split}
\end{eqnarray}
\end{subequations}
where we defined $k=N-|\beta_1|^2-|\beta_2|^2$.

The energy is minimized by four different solutions depending on whether the couplings exceed the critical coupling
\begin{eqnarray}
	\lambda_{c} = \frac{\sqrt{\omega\omega_0}}{2},
\end{eqnarray}

i)  When both  couplings are smaller than the critical coupling $\lambda\equiv\max(\lambda_1,\lambda_2)<\lambda_{c}$, the trivial solution 
\begin{eqnarray}
	\alpha=\beta_1=\beta_2=0
\end{eqnarray}
survives, corresponding to the normal state (N). 

ii) and iii) When either coupling becomes larger than the critical coupling but the two coulings are unequal $\lambda>\lambda_c$, $\lambda_1\neq\lambda_2$, the system is superradiant (S). For $\lambda_1>\lambda_2$ ($\lambda_2>\lambda_1$), only the $b_0$ and $b_1$ ($b_2$) modes are occupied macroscopically
\begin{subequations}\label{eq:closed_SR_expect}
\begin{eqnarray}
	\begin{split}
	\beta_1 = \pm|\beta|,\quad \beta_2=0,\quad \alpha=\pm i |\alpha|, \quad&& \lambda_1>\max(\lambda_{c},\lambda_2), \\
	\beta_1 = 0,\quad \beta_2=\pm i |\beta|,\quad \alpha=\pm |\alpha|,\quad&& \lambda_2>\max(\lambda_{c},\lambda_1),
	\end{split}
\end{eqnarray}
with
\begin{eqnarray}
	|\alpha|&=&\frac{\lambda\sqrt{N}}{\omega}\sqrt{1-\frac{\lambda_{c,\mathrm{cl}}^4}{\lambda^4}},\quad  |\beta|= \sqrt{\frac{N}{2}}\sqrt{1-\frac{\lambda_{c,\mathrm{cl}}^2}{\lambda^2}},
\end{eqnarray} 
and the $\mathbb{Z}_2$ symmetry of the  Hamiltonian is reflected by the choice of the sign prefactor.
\end{subequations}

 iv) Finally, when the two couplings are equal and above the critical coupling $\lambda_1=\lambda_2>\lambda_{c}$, the superradiant solution is a superposition involving all three modes
\begin{eqnarray}
	\beta_1 = |\beta|\cos\phi,\quad \beta_2=i|\beta|\sin\phi,\quad \alpha= |\alpha|e^{-i\left(\phi-\frac{\pi}{2}\right)},
\end{eqnarray}
where the $\mathrm{U}(1)$ symmetry is reflected by the choice of the phase $\phi$.
As expected from the form of the Hamiltonian, the coupling between the cavity and atomic fields exists between the real (imginary) part of the $b_1$ ($b_2$) field and the imaginary (real) part of the $a$ field.

After substituting the mean-field solutions into the HP Hamiltonian, 
we can now write down the HP Hamiltonian for the fluctuations around the normal (N) and superradiant (S) states. Without loss of generality, we only show the results for $\lambda_1\ge\lambda_2$, and choose $\phi=0$ for the case of equal couplings:
\begin{eqnarray}\label{eq:hamil_hp_normal_SR}
	H_\mathrm{N/S}&=&\omega c^\dagger c + \tilde{\omega}_0 (d_1^\dagger d_1+d_2^\dagger d_2) + \tilde{\Omega}_1 (d_1^\dagger+d_1)^2 + i \tilde{\lambda}_1 (c-c^\dagger)(d_1^\dagger + d_1)+ i \tilde{\lambda}_2 (c+c^\dagger)(d_2 - d_2^\dagger),
\end{eqnarray}
with
\begin{eqnarray}\label{eq:coef_hp_closed_sr}
\begin{split}
&\tilde{\lambda}_1= \frac{\sqrt{2}\lambda_1 \mu_1}{\sqrt{1+\mu_1}},\quad
\tilde{\lambda}_2= \frac{\lambda_2}{\sqrt{2}}\sqrt{1+\mu_1}\\
&\tilde{\Omega}_1= \frac{\lambda_1^2}{2\omega}\frac{(1-\mu_1)(3+\mu_1)}{1+\mu_1},\quad 
\tilde{\omega}_0 = \frac{\omega_0}{2}\left(1+\frac{1}{\mu_1} \right) \\
& \mu_1 = \begin{cases}
		1,\quad & \text{(N)} \\
		\lambda_{c}^2/\lambda_1^2.\quad &  \text{(S)}
		\end{cases}
\end{split}
\end{eqnarray}

\subsubsection{Inverted states}

Although the inverted states are excited states of the closed system and therefore do not appear in the phase diagram [cf. Fig.~1(b) of the main text], it is useful to find its HP Hamiltonian, and solve its stability. This can serve as a precursor for the analysis in the open system. 

The HP transformation for the inverted states has a slightly different procedure. Instead of Eq.~\eqref{eq:HP_mapping1}, we formally perform the following transformation
\begin{eqnarray}
	\begin{split}
	\Sigma_{01}^\dagger = \Sigma_{10}&\to \sqrt{N-b_0 ^\dagger b_0  - b_2^\dagger b_2}\,\,b_0, \\ 
	\Sigma_{02}^\dagger = \Sigma_{20} &\to b_2^\dagger b_0, \\
	\Sigma_{11}&\to N - b_0^\dagger b_0 - b_2^\dagger b_2,\\
	\Sigma_{22}&\to b_2^\dagger b_2.
	\end{split}
\end{eqnarray}
Similar to the previous case, we can split the new operators into their expectation values and fluctuation operators. We note that the expectation values $\langle a\rangle$, $\langle b_0\rangle$ and $\langle b_2\rangle$ are by definition of the inverted states $0$, $0$ and $\sqrt{N-N_1}e^{i\theta}$, respectively [cf. Eq.~(3) of the main text]:
\begin{eqnarray}
a = c,\quad 
b_0 = d_0, \quad 
b_2 = d_2 + \sqrt{N-N_1}e^{i\theta}.
\end{eqnarray}
Up to the terms scaling as $\mathcal{O}(1)$ in terms of $N$, the HP Hamiltonian describing the fluctuations around the inverted states (I) reads
\begin{subequations}\label{eq:hamil_hp_inverted}
\begin{eqnarray}
	H_\mathrm{I} &=& \omega c^\dagger c - \omega_0 d_0^\dagger d_0 + i\eta_1 (c-c^\dagger)(d_0^\dagger + d_0) +i\eta_2 (c+c^\dagger) (e^{i\theta}d_0^\dagger - e^{-i\theta}d_0),
\end{eqnarray}
where we have defined the notations
\begin{eqnarray}\label{eq:def_etas}
	\eta_1 = \lambda_1 \sqrt{\frac{N_1}{N}} ,\quad \eta_2=\lambda_2\sqrt{1-\frac{N_1}{N}}.
\end{eqnarray}
\end{subequations}

For equal couplings $\lambda_1=\lambda_2$, the $\mathrm{U}(1)$ symmetry present in the original Hamiltonian Eq.~\eqref{eq:hamil_suppmat} can also be reflected in the HP Hamiltonian for the inverted states. The HP Hamiltonian remains invariant under the transformation of the operators $(c,d_0)$ and the inverted state parameters $(N_1,\theta)$:
\begin{align}\label{eq:U1inHP}
	\begin{split}
	c&\mapsto ce^{i\phi},\\
	d_0&\mapsto d_0e^{i\chi},\\
	N_1&\mapsto N\sin^2\phi + N_1(\cos^2\phi-\sin^2\phi) - 2\sqrt{N_1(N-N_1)}\sin\phi\cos\phi\sin\theta,\\
	\theta&\mapsto \arctan\left(\frac{(2N_1-N)\sin2\phi + 2\sqrt{N_1(N-N_1)}\cos2\phi\sin\theta}{2\sqrt{N_1(N-N_1)}\cos\theta}\right),\\
	\chi&= \arctan\left(\frac{\sqrt{N-N_1}\tan\phi\cos\theta}{\sqrt{N_1}+\sqrt{N-N_1}\tan\phi\sin\theta}\right),
	\end{split}
\end{align}
where $\phi\in[0,2\pi)$ is a parameter associated to the $\mathrm{U}(1)$ symmetry.
Notably, this transformation is consistent with the generator $\mathcal{G}$ of the $\mathrm{U}(1)$ symmetry presented in the main text, and furthermore $N_1\sqrt{N-N_1}\cos\theta$ remains invariant under this mapping, as is consistent with the contours shown in Fig.~2(b) of the main text.

\subsection{Polaritonic excitation spectra and stability}

\begin{figure}
	\centering
	\includegraphics[width=\columnwidth]{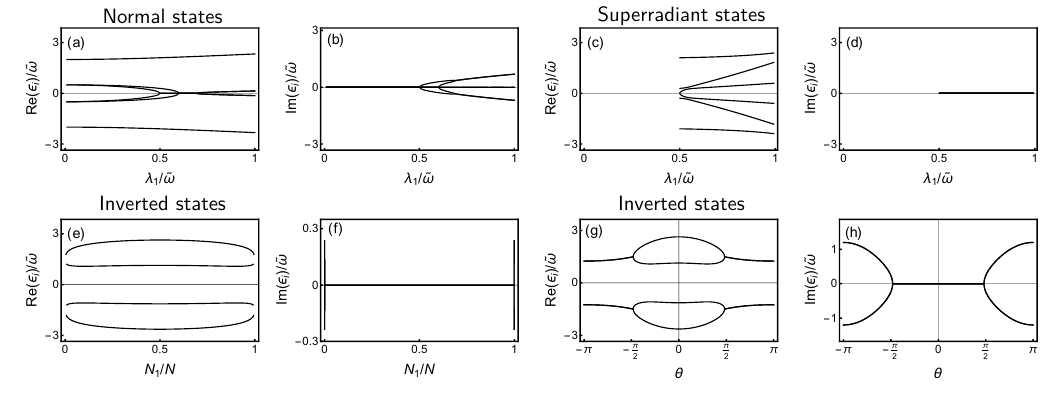}
	\caption{
	The (a,c,e,g) real and (b,d,f,h) imaginary parts of the excitation spectra for (a,b) the normal states, (c,d) the superradiant states, and (e-h) the inverted states. The cavity frequency and atomic frequency are chosen as $\omega=2\tilde{\omega}$ and $\omega_0=\tilde{\omega}/2$, respectively. (a-d) For the normal and superradiant states, we further choose $\lambda_2=\lambda_1/1.2$ and tune $\lambda_1$ from $0$ to $\tilde{\omega}$. (e-h) For the inverted states, we further choose $\lambda_1=\lambda_2=\tilde{\omega}$.  In panels (g-i), we fix $\theta=0$ and tune $N_1/N$ from $0$ to $1$. In panels (j-l), we fix $N_1=N/2$, and tune $\theta$ from $-\pi$ to $\pi$. 
	}
	\label{fig:closed_spectrum_fluc}
\end{figure}

We have obtained the HP Hamiltonian describing the fluctuations around the three states in interest. These are all quadratic, bosonic Hamiltonians, and can be solved exactly by the Hopfield-Bogoliubov transformation method~\cite{hopfield58,xiao09}. In this section, we first briefly recapitulate the procedure of this method, and then discuss the obtained spectra and fluctuations of the three states.

A quadratic Hamiltonian $H$ of $n$ bosonic modes can be generally expressed by~\footnote{In Sections~\ref{sec:closed} and \ref{sec:open} of this Supplementary Material, the complex conjugate, transpose, and Hermitian conjugate of a matrix $\mathbf{A}$ are denoted as $\mathbf{A}^*$, $\mathbf{A}^\mathrm{T}$, and $\mathbf{A}^\dagger$, respectively.}
\begin{subequations}
\begin{eqnarray}\label{eq:def_H_and_K}
	H = \underline{a}^\dagger \mathbf{H} \underline{a} + \frac{1}{2}\underline{a} \mathbf{K}\underline{a} +  \frac{1}{2}\underline{a}^\dagger \mathbf{K}^*\underline{a}^\dagger,
\end{eqnarray}
where the basis
\begin{eqnarray}
	\underline{a}=(a_1,a_2,\dots,a_n)^\mathrm{T}
\end{eqnarray}
\end{subequations}
is a vector of bosonic annihilation operators. $\mathbf{H} = \mathbf{H}^\dagger$ is Hermitian while $\mathbf{K} = \mathbf{K}^\mathrm{T}$ is symmetric, and both of them are $n$-dimensional. We can construct the corresponding $2n$-dimensional Hopfield-Bogoliubov matrix as~\footnote{In Sections~\ref{sec:closed} and \ref{sec:open} of this Supplementary Material, the bold typeface $\mathbf{A}$ and the single underline $\underline{a}$ are used to denote $n\times n$ matrices and $n$-dimensional vectors, respectively; whereas the caligraphic typeface $\mathcal{A}$ and the double underlines $\underline{\underline{a}}$  are used to denote $2n\times 2n$ matrices and $2n$-dimensional vectors, respectively.}
\begin{subequations}
\begin{eqnarray}\label{eq:def_HB_matrix}
	\mathcal{D} = \begin{pmatrix}
		\mathbf{H} &\mathbf{K} \\ -\mathbf{K}^\dagger & -\mathbf{H}^\mathrm{T}
	\end{pmatrix}
\end{eqnarray}
under the basis
\begin{eqnarray}
	\underline{\underline{a}}=(\underline{a},\underline{a}^{\dagger\mathrm{T}})^\mathrm{T}.
\end{eqnarray}
\end{subequations}

For the normal and superradiant state, we choose $\underline{a}=(c,d_1,d_2)^\mathrm{T}$, and thus for $\lambda_1\ge\lambda_2$, the two matrices read
\begin{eqnarray}\label{eq:H_and_K_NO_SR}
\mathbf{H}_{\mathrm{N/S}}= \begin{pmatrix}
	\omega & -i\tilde{\lambda}_1 &i\tilde{\lambda}_2\\
	i\tilde{\lambda}_1 &\tilde{\omega}_0 + 2\tilde{\Omega}_1 & 0\\
	-i\tilde{\lambda}_2 &0 &\tilde{\omega}_0
\end{pmatrix},\quad
\mathbf{K}_{\mathrm{N/S}} = \begin{pmatrix}
	0 & -i\tilde{\lambda}_1 &-i\tilde{\lambda}_2\\
	-i\tilde{\lambda}_1 & 2\tilde{\Omega}_1 & 0\\
	-i\tilde{\lambda}_2 &0 &0
\end{pmatrix},
\end{eqnarray}
with the parameters defined in Eq.~\eqref{eq:coef_hp_closed_sr}. For the inverted state, by choosing $\underline{a}=(c,d_0)^\mathrm{T}$, the matrices read
\begin{eqnarray}\label{eq:H_and_K_for_inverted}
	\mathbf{H}_{\mathrm{I}}= \begin{pmatrix}
		\omega & -i\lambda_{+-}  \\
		i\lambda_{++} & -\omega_0 \\
	\end{pmatrix},\quad
	\mathbf{K}_{\mathrm{I}} = \begin{pmatrix}
		0 & -i\lambda_{-+} \\
		-i\lambda_{-+} & 0 \\
	\end{pmatrix},
\end{eqnarray}
where 
\begin{eqnarray}\label{eq:def_lambda_pm}
	\lambda_{\sigma\rho} \equiv \eta_1 +\sigma \eta_2 e^{\rho i \theta},\quad \sigma=\pm1,\,\quad \rho=\pm1,
\end{eqnarray}
and $\eta_1$ and $\eta_2$ are defined in Eq.~\eqref{eq:def_etas}.

Diagonalization of the Hopfield-Bogoliubov matrix gives the excitation spectrum for the fluctuations around the respective states~\cite{xiao09}. For a quadratic Hamiltonian with $n$ bosonic operators,
it involves a transformation matrix $\mathbf{T}$ such that
\begin{subequations}
\begin{eqnarray}
	\mathcal{T}^{-1}\mathcal{D}\mathcal{T} &=&\mathbf{E}\otimes\tau_3, \\
	\mathbf{E} &=& \mathrm{diag}(\epsilon_1,\epsilon_2,\dots,\epsilon_n),
\end{eqnarray}
\end{subequations}
where $\epsilon_i\ge0$ constitute the excitation spectrum of the fluctuations, and $\tau_3$ denotes the third Pauli matrix. The transformation matrix follows an unusual normalization $\mathcal{T}^\dagger\mathcal{I}_- \mathcal{T} = \mathcal{I}_-$, with $\mathcal{I}_-=\mathbf{1}_{n\times n}\otimes\tau_3$. Generally in our system, the transformation matrix has to be determined numerically.
Once $\mathcal{T}$ is determined, we can transform into the eigenbasis of the Hopfield-Bogoliubov matrix $\underline{\underline{\tilde{a}}}=(\underline{\tilde{a}},\underline{\tilde{a}}^{\dagger\mathrm{T}})^\mathrm{T}$ by $\underline{\underline{\tilde{a}}} = \mathcal{T}^{-1}\underline{\underline{a}}$. In the new basis, the Hamiltonian is diagonalized and can be rewritten as
\begin{eqnarray}
	H = \underline{\tilde{a}}^\dagger \mathbf{E} \underline{\tilde{a}}.
\end{eqnarray}
Notice that in the diagonalized representation, the Hamiltonian no longer contains terms consisting of double creation/annihilation operators like $\tilde{a}_i^\dagger\tilde{a}_i^\dagger$ and $\tilde{a}_i\tilde{a}_i$, i.e., $\tilde{\mathbf{K}}=\mathbf{0}$.

The excitations on top of the eigenstates behave like polaritons~\cite{emary03}, and their energy is in the order of $\mathcal{O}(1)$ in terms of $N$. In Fig.~\ref{fig:closed_spectrum_fluc}, we show exemplary polaritonic excitation spectra of the different states. When the system has a completely real spectrum, the system is physical and stable. The normal state is stable for $\lambda<\lambda_{c}$, while the superradiant state is stable for $\lambda>\lambda_{c}$. This result is summarized as Fig.~1(b) of the main text. Apart from the ground state, we also briefly discuss the stability of the excited inverted states. They are stable in the $N_1$\nobreakdash-$\theta$ phase space within the region
\begin{eqnarray}\label{eq:inverted_stable_boundary_closed}
\omega \omega_0(\eta_1^2+\eta_2^2) - \eta_1 \eta_2 (\omega^2+\omega_0^2)\cos\theta = \frac{1}{16}(\omega^2-\omega_0^2)^2,
\end{eqnarray}
where $\eta_1$ and $\eta_2$ are defined in Eq.~\eqref{eq:def_etas}.

Besides the excitation spectrum, the expectation values of the closed-system cavity and atomic fluctuations can also be obtained. Generally, the fluctuation of the $a_i$ mode $\langle a_i^\dagger a_i\rangle$ is given by
\begin{eqnarray}\label{eq:closed_fluctuations}
	\begin{split}
	\langle a_i^\dagger a_i\rangle &= \langle\underline{\tilde{a}}^\dagger \mathcal{T}^{-1} \mathcal{J}_{i}
	\mathcal{T} \underline{\tilde{a}}\rangle = \sum_{k=n+1}^{2n}\vert\mathcal{T}_{i,k}\vert^2 \\
	\mathcal{J}_{i} &= \mathrm{diag}(\underbrace{0,\dots,0}_{(i-1)\text{ times 0}},\underbrace{1}_{i\text{th}},\underbrace{0,\dots,0}_{(2n-i)\text{ times 0}}),
	\end{split}
\end{eqnarray}
where we have used the fact that $\langle \tilde{a}_i \tilde{a}^\dagger_j\rangle = \delta_{i,j}$ and $\langle \tilde{a}_i^\dagger \tilde{a}_j\rangle = 0$ in the diagonalized basis. Other correlations like $\langle a_i^\dagger a_j\rangle$ and $\langle a_ia_j\rangle$ can be obtained in a similar manner. This result will later be compared to the fluctuations in the open system.

\section{Steady states and their stability in the open system}\label{sec:open}

In this section, we further perform a similar analysis to investigate the fixed-point states and their stability in the open system.
In the open system, the dissipation of the cavity can be captured by the Lindblad operator $L=\sqrt{\kappa}a$. In this case, the equation of motion for the density matrix is given by Eq.~(4) of the main text, while the equation of motion for an operator $O$ is given by~\cite{gardiner85}
\begin{eqnarray}\label{eq:equation_of_motion}
	\frac{d}{dt} O = -\frac{i}{\hbar}[O,H] -\kappa \left\{ [O,a^\dagger]a - a^\dagger [O,a]  \right\}.
\end{eqnarray}

In the following, we first find the HP Hamiltonians for the different states in the open system. Particularly, we observe that the normal and inverted states share the same HP Hamiltonians as in the closed system, while the superradiant state has a different one. We then perform a third quantization analysis upon the obtained HP Hamiltonians, and calculate the stability of the states. Furthermore, we obtain the cavity fluctuations by solving equations of motion based on the HP Hamiltonians. Finally, we derive the system dynamics arising from the Liouvillian equations of motion, which is used to probe the system stability in the main text.

\subsection{Superradiant steady state and its Holstein-Primakoff Hamiltonian}

Similar to the procedure in the closed system, we calculate the HP Hamiltonian for the fixed-point states. This remains unchanged for the normal and inverted states as Eq.~\eqref{eq:hamil_hp_normal_SR} and Eq.~\eqref{eq:hamil_hp_inverted}, respectively. This is because they have the same expectation values for the cavity and atomic fields in closed and open systems. Nevertheless, the solution for the superradiant state is significantly affected by dissipation. Therefore, we first solve the HP Hamiltonian for the superradiant state. 

As a fixed point, the time derivatives of all the expectation values vanish in the superradiant state. The equations $d\langle \Sigma_{11}\rangle /dt=0$, $d\langle \Sigma_{22}\rangle /dt=0$ and $d\langle a\rangle/dt = 0$ lead to
\begin{eqnarray}
	\langle \Sigma_{01}\rangle\in\mathbb{R},\quad \langle \Sigma_{02}\rangle\in\mathbb{I}, \quad 	\langle a \rangle = \frac{2i\left(\lambda_1\langle \Sigma_{01}\rangle-\lambda_2\langle \Sigma_{02}\rangle\right)}{\sqrt{N}(\omega-i\kappa)},
\end{eqnarray}
respectively. Treating the cavity field as a coherent light $a\to\langle a\rangle$ and substituting the above relation into the Hamiltonian Eq.~\eqref{eq:hamil_suppmat}, we obtain an effective Hamiltonian
\begin{eqnarray}
	H  = \omega_0 (\Sigma_{11}+\Sigma_{22}) + \Lambda_1 (\Sigma_{01}+\Sigma_{10}) + i\Lambda_2 (\Sigma_{02}-\Sigma_{20}),
\end{eqnarray}
where the coefficients $\Lambda_1$  and $\Lambda_2$ are defined by the expectation values of the atomic field
\begin{eqnarray}\label{eq:open_SR_expect1}
	\Lambda_1 =\frac{4\lambda_1\left(-\omega\lambda_1 \langle \Sigma_{01} \rangle + i\kappa\lambda_2 \langle \Sigma_{02}\rangle  \right)}{N(\omega^2+\kappa^2)}\in\mathbb{R},\quad\quad
	\Lambda_2 = \frac{4\lambda_2\left(-\kappa\lambda_1 \langle \Sigma_{01}\rangle - i\omega\lambda_2 \langle \Sigma_{02}\rangle \right)}{N(\omega^2+\kappa^2)}\in\mathbb{R}.
\end{eqnarray}
Once these two coefficients are fixed, we can solve the Hamiltonian and find  the energy of the lowest-energy steady state 
\begin{eqnarray}
	E = \frac{N}{2}\left(\omega_0-\sqrt{4\Lambda_1^2+4\Lambda_2^2+\omega_0^2}\right)
\end{eqnarray}
as well as the expectation values of $\Sigma_{01}$ and $\Sigma_{02}$:
\begin{eqnarray}\label{eq:open_SR_expect2}
	\langle \Sigma_{01}\rangle = -\dfrac{N\Lambda_1}{\sqrt{4\Lambda_1^2+4\Lambda_2^2+\omega_0^2}},\quad\quad
	\langle \Sigma_{02}\rangle = \dfrac{iN\Lambda_2}{\sqrt{4\Lambda_1^2+4\Lambda_2^2+\omega_0^2}}.
\end{eqnarray}
Combining the two sets of equations Eqs.~\eqref{eq:open_SR_expect1} and ~\eqref{eq:open_SR_expect2}, we can solve $\Lambda_1$ and $\Lambda_2$ self-consistently.

\begin{subequations}\label{eq:open_SR_expect3}
\begin{eqnarray}
\Lambda_1 &=&\pm
\left(\frac{\kappa\lambda_1\lambda_2}{\omega^2+\kappa^2}\right) \sqrt{\frac{4\left[\omega(\lambda_1^2+\lambda_2^2)+\sqrt{\omega^2(\lambda_1^2-\lambda_2^2)^2-4\kappa^2\lambda_1^2\lambda_2^2}\right]^2 - \omega_0^2(\omega^2+\kappa^2)^2}
	{2\omega(\lambda_1^2-\lambda_2^2)\left[\omega(\lambda_1^2-\lambda_2^2)-\sqrt{\omega^2(\lambda_1^2-\lambda_2^2)^2-4\kappa^2\lambda_1^2\lambda_2^2}\right]}},\\
\Lambda_2 &=&\pm \mathrm{sgn}(\lambda_1-\lambda_2)
\left(\frac{\kappa\lambda_1\lambda_2}{\omega^2+\kappa^2}\right) \sqrt{\frac{4\left[\omega(\lambda_1^2+\lambda_2^2)+\sqrt{\omega^2(\lambda_1^2-\lambda_2^2)^2-4\kappa^2\lambda_1^2\lambda_2^2}\right]^2 -\omega_0^2(\omega^2+\kappa^2)^2}
	{2\omega(\lambda_1^2-\lambda_2^2)\left[\omega(\lambda_1^2-\lambda_2^2)+\sqrt{\omega^2(\lambda_1^2-\lambda_2^2)^2-4\kappa^2\lambda_1^2\lambda_2^2}\right]}}.
\end{eqnarray}
\end{subequations}
We note that the freedom in the choice of the sign prefactor reflects the two $\mathbb{Z}_2$ configurations. Since a superradiant state no longer exists for equal couplings $\lambda_1=\lambda_2$, the $\mathrm{U}(1)$ symmetry is no longer reflected in the fixed point solution. The suppression of the superradiance along the $\mathrm{U}(1)$ symmetry line is also observed in the interpolating Dicke--Tavis-Cummings model~\cite{soriente18}.

The superradiant state is a fixed point of the Liouvillian only when $\Lambda_1$ and $\Lambda_2$ are real. The phase boundaries can thus be solved and categorized into two distinctive sections,
\begin{subequations}\label{eq:normal_sr_boundary_open}
\begin{eqnarray}
\lambda_1 =  \lambda_2\left(\pm\frac{\kappa}{\omega}+\sqrt{\dfrac{\kappa^2}{\omega^2}+1}\right),\quad&& \lambda_1+\lambda_2>\lambda_{\mathrm{tr},1}+\lambda_{\mathrm{tr},2} 
\label{eq:normal_sr_boundary_open_a}\\
\lambda_1 = \dfrac{1}{2}\sqrt{\dfrac{\omega_0 [(\omega^2+\kappa^2)\omega_0-4\lambda_2^2\omega]}{(\omega\omega_0 - 4\lambda_2^2)}}, \quad&& \lambda_1+\lambda_2<\lambda_{\mathrm{tr},1}+\lambda_{\mathrm{tr},2}.
\label{eq:normal_sr_boundary_open_b}
\end{eqnarray}
\end{subequations}
The first kind includes two first-order lines where discontinuity in cavity expectation value $\langle a\rangle$ can be observed; while the second kind includes two second-order curves where divergence in cavity fluctuation $\langle a^\dagger a\rangle$ is seen. They intersect at the two tricritical points $(\lambda_{\mathrm{tr},1},\lambda_{\mathrm{tr},2})$ and $(\lambda_{\mathrm{tr},2},\lambda_{\mathrm{tr},1})$, with
\begin{eqnarray}
	\lambda_{\mathrm{tr},1,2} = \frac{1}{2}\sqrt{\frac{\omega\omega_0\sqrt{\omega^2+\kappa^2}}{\sqrt{\omega^2+\kappa^2}\pm\kappa}}.
\end{eqnarray}

Once the expectation values of the cavity and atomic fields in the superradiant state are solved, we can further write down the HP Hamiltonian also for the open system. This can be derived in a similar manner as the one presented in Ref.~\cite{dimer07}. The mean-field collective expectations of the atomic fields in the original and HP representations are related through $\langle \Sigma_{0\nu}\rangle = \beta_\nu \sqrt{N-|\beta_1|^2-|\beta_2|^2}$ for $\nu=1,2$, which, in combination with Eqs.~\eqref{eq:open_SR_expect2}, leads to
\begin{eqnarray}\label{eq:open_SR_expect_final}
	\begin{split}
	\beta_1 &= \frac{-\sqrt{2N}\Lambda_1}{\sqrt{\omega_0+\sqrt{4\Lambda_1^2+4\Lambda_2^2+\omega_0^2}}} \in\mathbb{R}, \\
	\beta_2 &= \frac{i\sqrt{2N}\Lambda_2}{\sqrt{\omega_0+\sqrt{4\Lambda_1^2+4\Lambda_2^2+\omega_0^2}}} \in\mathbb{I}.
	\end{split}
\end{eqnarray}
Using the obtained $\beta_{1,2}$ from Eq.~\eqref{eq:open_SR_expect_final} instead of the results from Eq.~\eqref{eq:closed_SR_expect}, we can calculate the HP Hamiltonian, and collect the terms up to the order of $\mathcal{O}(1)$ in terms of $N$:
\begin{eqnarray}
	\begin{split}
	H_\mathrm{S,open} =& \,\omega c^\dagger c + \tilde{\omega}_0 d_1^\dagger d_1 + \tilde{\omega}_0 d_2^\dagger d_2 \\
	& + i\tilde{\lambda}_1 (c-c^\dagger)(d_1+d_1^\dagger) +i\tilde{\lambda}_2 (c+c^\dagger) (d_2-d_2^\dagger)
	+ \tilde{\eta}_1 (c+c^\dagger)(d_1+d_1^\dagger) + \tilde{\eta}_2 (c-c^\dagger) (d_2-d_2^\dagger) \\
	&+\tilde{\Omega}_1 (d_1+d_1^\dagger)^2 + \tilde{\Omega}_2 (d_2-d_2^\dagger)^2
	+ i\tilde{K} (d_1+d_1^\dagger)(d_2-d_2^\dagger).
	\end{split}
\end{eqnarray}
The coefficients are all real numbers and are given below:
\begin{eqnarray}
	\begin{split}
		\tilde{\omega}_0
		&= \omega_0 +
		\frac{4\omega k(\lambda_1^2|\tilde{\beta}_1|^2+\lambda_2^2|\tilde{\beta}_2|^2)}{N(\omega^2+\kappa^2)}, \\
		\tilde{\lambda}_{\mu} &= \frac{\lambda_{\mu}\sqrt{k}}{\sqrt{N}} \left(1-|\tilde{\beta}_{\mu}|^2\right), \\
		\tilde{\eta}_{\mu} &= i\sigma_{\mu} \frac{\lambda_{\bar{\mu}}\sqrt{k}}{\sqrt{N}} \tilde{\beta}_1\tilde{\beta}_2, \\
		\tilde{\Omega}_{\mu} &= \frac{2k}{N(\omega^2+\kappa^2)}\left[\frac{\omega}{2}(\lambda_1^2|\tilde{\beta}_1|^2+\lambda_2^2|\tilde{\beta}_2|^2)|\tilde{\beta}_{\mu}|^2+\omega\lambda_{\mu}^2|\tilde{\beta}_{\mu}|^2+i\sigma_{\mu}\kappa\lambda_1\lambda_2\tilde{\beta}_1\tilde{\beta}_2\right], \\
		\tilde{K} &= \frac{-2k}{N(\omega^2+\kappa^2)}\left[(-i\omega\lambda_1^2\tilde{\beta}_1\tilde{\beta}_2+\kappa\lambda_1\lambda_2|\tilde{\beta}_2|^2)(1+|\tilde{\beta}_1|^2)
		+(i\omega\lambda_2^2\tilde{\beta}_1\tilde{\beta}_2+\kappa\lambda_1\lambda_2|\tilde{\beta}_1|^2)(1+|\tilde{\beta}_2|^2)\right].
	\end{split}
\end{eqnarray}
 For clarity, we have used the notations $\tilde{\beta}_{1,2} = \beta_{1,2}/\sqrt{k}$, $k=N-|\beta_1|^2-|\beta_2|^2$, $\sigma_1=-1$, $\sigma_2=1$, $\bar{1}=2$ and $\bar{2}=1$.

With this HP Hamiltonian $H_\mathrm{SR,open}$ for the superradiant states in the open system, we can rewrite the $\mathbf{H}$ and $\mathbf{K}$ matrices 
\begin{eqnarray}\label{eq:H_and_K_NO_SR_open}
	&&\mathbf{H}_\mathrm{S,open} = 
	\begin{pmatrix}
		\omega & -i\tilde{\lambda}_1 + \tilde{\eta}_1 &i\tilde{\lambda}_2 - \tilde{\eta}_2\\
		i\tilde{\lambda}_1 + \tilde{\eta}_1 &\tilde{\omega}_0+2\tilde{\Omega}_1 & i\tilde{K}\\
		-i\tilde{\lambda}_2 - \tilde{\eta}_2 & -i\tilde{K} &\tilde{\omega}_0+2\tilde{\Omega}_2
	\end{pmatrix}, \quad
	\mathbf{K}_\mathrm{S,open} = \begin{pmatrix}
		0 & -i\tilde{\lambda}_1+ \tilde{\eta}_1 &-i\tilde{\lambda}_2+ \tilde{\eta}_2\\
		-i\tilde{\lambda}_1 + \tilde{\eta}_1 & 2\tilde{\Omega}_1 & i\tilde{K}\\
		-i\tilde{\lambda}_2+ \tilde{\eta}_2 &i\tilde{K} &2\tilde{\Omega}_2
	\end{pmatrix}.
\end{eqnarray}

\subsection{Stability of steady states}

\begin{figure}[t]
	\centering
	\includegraphics[width=0.9\columnwidth]{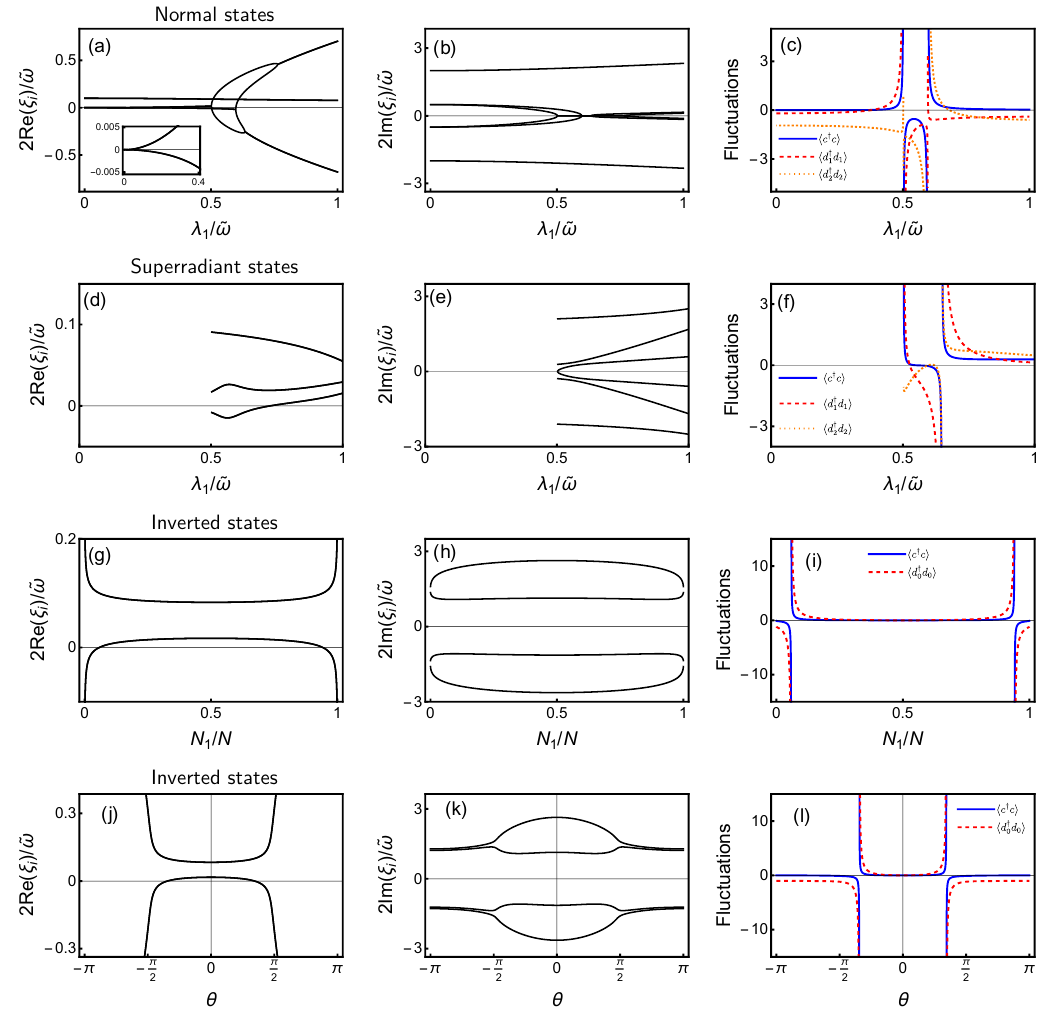}
	\caption{
		The (a,d,g,j) real and (b,e,h,k) imaginary parts of the rapidities, and (c,f,i,l) the cavity and atomic fluctuations for (a-c) the normal states, (d-f) the superradiant states, and (g-l) the inverted states. 
		The parameters of the system are $\omega=2\tilde{\omega}$, $\omega_0=\tilde{\omega}/2$, $\kappa=0.1\tilde{\omega}$. 
		(a-f) For the normal and superradiant states, we further choose $\lambda_2=\lambda_1/1.2$ and tune $\lambda_1$ from $0$ to $\tilde{\omega}$. (g-l) For the inverted state, we further fix $\lambda_1=\lambda_2=\tilde{\omega}$. In panels (g-i), we fix $\theta=0$ and tune $N_1/N$ from $0$ to $1$. In panels (j-l), we fix $N_1=N/2$, and tune $\theta$ from $-\pi$ to $\pi$. 
	}
	\label{fig:open_spectrum_fluc}
\end{figure}

The fixed-point states of the open system become steady states only when they are stable, since the system will be driven away from unstable fixed-point states by fluctuations. The stability of the fixed-point states can be found by third quantization analysis~\cite{prosen08,prosen10} based on the HP Hamiltonian. Third quantization solves the eigenvalues of a Liouvillian exactly by quantizing the density matrix. Similar to the Hopfield-Bogoliubov transformation method, it can be applied on a quadratic Hamiltonian in the form of Eq.~\eqref{eq:def_H_and_K} with a linear Lindbladian in the form of
\begin{eqnarray}
	L = \underline{\ell}_1 \cdot\underline{a} + \underline{\ell}_2\cdot \underline{a}^\dagger.
\end{eqnarray}
We briefly introduce the technical procedures relevant to the investigation of system stability.

The dynamical behavior of the system can then be revealed by the shape matrix of the Liouvillian~\cite{prosen08}
\begin{subequations}
	\begin{eqnarray}
		\mathcal{X} = \frac{1}{2}
		\begin{pmatrix}
			i\mathbf{H}^*-\mathbf{N}^*+\mathbf{M} & -i\mathbf{K}-\mathbf{L}+\mathbf{L}^\mathrm{T} \\
			i\mathbf{K}^*-\mathbf{L}^*+\mathbf{L}^\dagger & -i\mathbf{H}-\mathbf{N}+\mathbf{M}^*
		\end{pmatrix},
	\end{eqnarray}
where $\mathbf{H}$ and $\mathbf{K}$ are defined in Eq.~\eqref{eq:def_H_and_K}. For the normal, superradiant and inverted states of our system, the $\mathbf{H}$ and $\mathbf{K}$ matrices are given by Eq.~\eqref{eq:H_and_K_NO_SR}, Eq.~\eqref{eq:H_and_K_NO_SR_open}, and Eq.~\eqref{eq:H_and_K_for_inverted}, respectively.
	
The other three matrices are defined by the dissipation channel,
	\begin{eqnarray}
		\mathbf{M} = \underline{\ell}_1\otimes \underline{\ell}_1^*, \quad \mathbf{N} = \underline{\ell}_2\otimes \underline{\ell}_2^*, \quad \mathbf{L} = \underline{\ell}_1\otimes \underline{\ell}_2^* .
	\end{eqnarray}
\end{subequations}
For our system the Lindbladian is $L=\sqrt{\kappa}c$ in the HP representation. Therefore, for the normal and superradiant states where the basis is chosen as $\underline{a}=(c,d_1,d_2)^\mathrm{T}$, we have $\underline{\ell}_1=(\sqrt{\kappa},0,0)^\mathrm{T}$ and $\underline{\ell}_2=(0,0,0)^\mathrm{T}$, yielding 
\begin{equation}
	\mathbf{M}_{\mathrm{N/S}} = \mathrm{diag}(\kappa,0,0),\quad
	\mathbf{N}_{\mathrm{N/S}} = \mathbf{L}_{\mathrm{N/S}} = \mathbf{0}_{3\times3};
\end{equation}
while for the inverted states where the basis is chosen as $\underline{a}=(c,d_0)^\mathrm{T}$, we have $\underline{\ell}_1=(\sqrt{\kappa},0)^\mathrm{T}$ and $\underline{\ell}_2=(0,0)^\mathrm{T}$, yielding 
\begin{equation}
	\mathbf{M}_{\mathrm{I}} =  \mathrm{diag}(\kappa,0),\quad
	\mathbf{N}_{\mathrm{I}} = \mathbf{L}_{\mathrm{I}} = \mathbf{0}_{2\times2}.
\end{equation}
In the absence of dissipation $L=0$, the matrix $\mathcal{X}$ is essentially the Hopfield-Bogoliubov matrix [Eq.~\eqref{eq:def_HB_matrix}] up to a prefactor and a Hermitian conjugation $\mathcal{D} = 2i\mathcal{X}^\dagger(L=0)$.

The eigenvectors of $\mathcal{X}$ describe the polaritonic excitation states on top of the steady states, whereas the eigenvalues, called rapidities $\xi_i$, are negatively related to the eigenvalues of the Liouvillian. For a stable steady state, the real parts of all the rapidities are nonnegative, i.e., $\min\Re\xi_i\ge0$. Like the polaritonic excitation spectra in the close system, these rapidities also scale as $\mathcal{O}(1)$ in terms of $N$.
Examples of rapidities for the normal, superradiant, and inverted states are shown in Fig.~\ref{fig:open_spectrum_fluc}(a,b,d,e,g,h,j,k). As explained in the main text, the real part of the rapidities determine the stability of the corresponding fixed-point states. By solving the rapidities for different parameters, we can obtain the stability diagrams for the three states as summarized in Fig.~1(c) of the main text. On the other hand, the imaginary part of the rapidities describe the energy of the polaritonic excitation, and is thus related to the oscillation frequency of the system evolution.

\subsection{Fluctuations around the steady states}

\begin{figure}[ht]
	\centering
	\begin{minipage}{\columnwidth}
	\includegraphics[width=0.75\columnwidth]{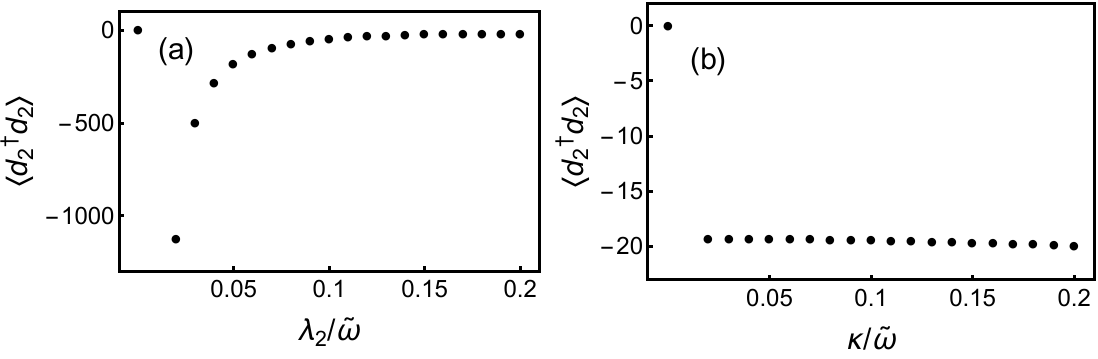}
	\caption{
		The atomic fluctuations of the $b_2$ mode, $\langle d_2^\dagger d_2\rangle$, as functions of (a) $\lambda_2$ and (b) $\kappa$. Parameters are chosen as $\omega=2\tilde{\omega}$, $\omega_0=\tilde{\omega}/2$, $\lambda_1=0.3\tilde{\omega}$, and $\kappa=0.1$ in panel (a) and $\lambda_2=0.2$ in panel (b). Divergence and discontinuity can be seen at $\lambda_2=0$ and $\kappa=0$, respectively. The fluctuations are calculated using Eq.~\eqref{eq:atomic_fluc_open} for the open system ($\kappa\neq0$), and Eq.~\eqref{eq:closed_fluctuations} for the closed system ($\kappa=0$).
	}
	\label{fig:open_fluc_limits}
	\end{minipage}
	\begin{minipage}{\columnwidth}
	\includegraphics[width=0.75\columnwidth]{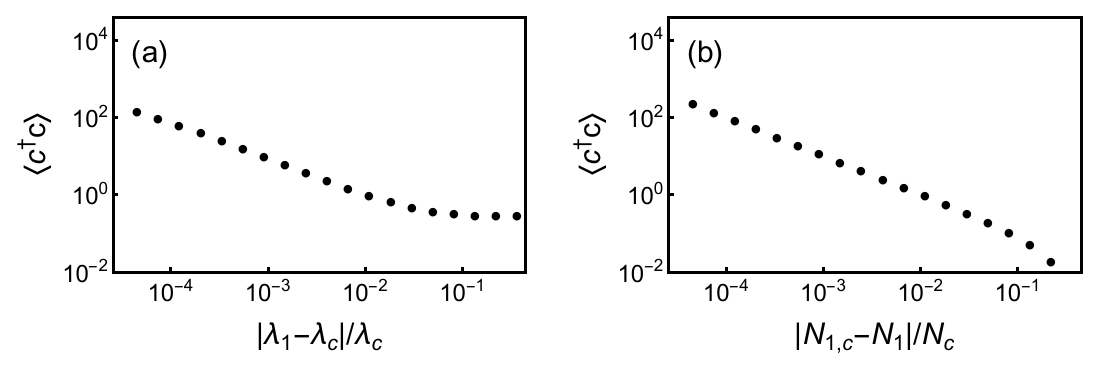}
	\caption{
		The cavity fluctuations plotted in logarithmic scale in the vicinity of stability boundaries for (a) the superradiant states and (b) the inverted states.
		In panel (a), the parameters are chosen as $\omega=2\tilde{\omega}$, $\omega_0=\tilde{\omega}/2$, $\kappa=0.1\tilde{\omega}$, $\lambda_1=1.2\lambda_2$. The stability boundary is numerically determined as $\lambda_c\approx0.648\tilde{\omega}$. 
		In panel (b)), parameters are chosen as $\omega=2\tilde{\omega}$, $\omega_0=\tilde{\omega}/2$, $\kappa=0.1\tilde{\omega}$, $\lambda_1=\lambda_2=\tilde{\omega}$, and $\theta=0$. The stability boundary is $N_{1,c}=\left(\frac{1}{2}+\frac{\sqrt{35369}}{426}\right)N\approx0.94N$. In both cases, the stability boundary is approached from the stable side.
	}
	\label{fig:fluc_exponent}
\end{minipage}

\end{figure}

\begin{figure}
	\centering
	\includegraphics[width=0.6\columnwidth]{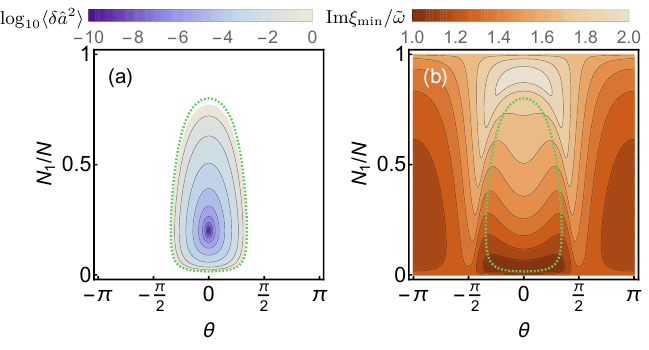}
	\caption{
		(a) Cavity fluctuations $\langle \delta a^2\rangle = \langle c^\dagger c\rangle$ and (b) imaginary part of the rapidity $\vert\Im\xi_\mathrm{min}\vert$ of the inverted states when the two couplings are unequal. Parameters are taken as $\omega=2\tilde{\omega}$, $\omega_0=0.5\tilde{\omega}$, $\kappa=0.1\tilde{\omega}$, $\lambda_1=2\tilde{\omega}$ and $\lambda_2=\tilde{\omega}$. The green dotted line indicates the stability boundary and is reproduced from Fig.~2(a) of the main text.
	}
	\label{fig:non_equal_coupling_spectra}
\end{figure}

The fluctuations around the steady states are described by the HP Hamiltonian. To solve their expectation values, we can apply the equations of motion Eq.~\eqref{eq:equation_of_motion} based on the HP Hamiltonian~\cite{dimer07,baric12,nagy10,soriente18}. Since the cavity fluctuations are solved in a fixed-point state, we require that all correlation operators have vanishing time derivatives. For example, the equations of motion for the inverted states are given by ($\lambda_{\pm\pm}$ defined in Eq.~\eqref{eq:def_lambda_pm})
\begin{eqnarray}
	\begin{split}
	\frac{d}{dt} \langle cc\rangle &= -(2i\omega+2\kappa) \langle cc\rangle - 2 \lambda_{-+}\langle cd_0^\dagger \rangle - 2\lambda_{+-}\langle cd_0\rangle=0,\\
	\frac{d}{dt} \langle c^\dagger c\rangle &= -\lambda_{-+}\langle c^\dagger d_0^\dagger\rangle - \lambda_{+-} \langle c^\dagger d_0\rangle - \lambda_{--} \langle cd_0\rangle - \lambda_{++} \langle cd_0^\dagger\rangle   -2\kappa \langle c^\dagger c\rangle=0,  \\
	\frac{d}{dt} \langle d_0d_0\rangle &= 2i\omega_0 \langle d_0d_0\rangle + 2\lambda_{++} \langle cd_0\rangle - 2\lambda_{-+} \langle c^\dagger d\rangle=0, \\
	\frac{d}{dt} \langle d_0^\dagger d_0\rangle &= \lambda_{++} \langle cd_0^\dagger \rangle - \lambda_{-+} \langle c^\dagger d_0^\dagger\rangle + \lambda_{+-} \langle c^\dagger d_0\rangle - \lambda_{--} \langle cd_0\rangle=0, \\
	\frac{d}{dt} \langle cd_0\rangle &= -i(-\omega_0+\omega-i\kappa) \langle cd_0\rangle-\lambda_{-+}\langle d_0^\dagger d_0\rangle - \lambda_{+-} \langle d_0d_0\rangle + \lambda_{++} \langle cc\rangle - \lambda_{-+} \langle c^\dagger c\rangle - \lambda_{-+}=0, \\	
	\frac{d}{dt} \langle c^\dagger d_0\rangle &= -i(-\omega_0-\omega-i\kappa)\langle c^\dagger d_0\rangle - \lambda_{--} \langle d_0d_0\rangle - \lambda_{++} \langle d_0^\dagger d_0\rangle + \lambda_{++} \langle c^\dagger c\rangle -\lambda_{-+} \langle c^\dagger c^\dagger\rangle=0,
	\end{split}
\end{eqnarray}
and their complex conjugated versions, giving 10 equations in total. 
These equations can be solved simultaneously, yielding the cavity fluctuation $\langle c^\dagger c\rangle = \langle \delta a^2\rangle = \langle a^\dagger a\rangle - |\langle a\rangle|^2$ and atomic fluctuations $\langle d_\mu^\dagger d_\mu\rangle = \langle b_\mu^\dagger b_\mu\rangle - |\langle b_\mu\rangle|^2$ around the steady states.
As an example, the cavity fluctuations around the inverted states are given by Eq.~(7) of the main text. We note that this set of equations is also valid in the closed system by setting $\kappa=0$. However, the equations then become underdetermined, having the closed-system fluctuations Eq.~\eqref{eq:closed_fluctuations} as one of the solutions.

Likewise, the same analysis on the normal state based on Eq.~\eqref{eq:hamil_hp_normal_SR} gives a set of 21 equations.
The singularity of the system at $\lambda_1=0$, $\lambda_2=0$, or $\kappa=0$ manifests itself in this set of equations. When $\lambda_1=0$ or $\lambda_2=0$, the number of independent equations is reduced from 21 to 20, and such mathematical singularity is also captured in fluctuations. Specifically, the atomic fluctuation of the $b_2$ mode,
\begin{eqnarray}\label{eq:atomic_fluc_open}
	\langle d_2^\dagger d_2\rangle = \frac{\lambda_2^2-\lambda_1^2}{8\lambda_2^2}\left(\frac{\omega}{\omega_0}+\frac{\omega^2(\omega\omega_0-4\lambda_1^2)+\kappa^2(4\lambda_2^2+\omega\omega_0)}{16\lambda_1^2\lambda_2^2-4(\lambda_1^2+\lambda_2^2)\omega\omega_0+(\omega^2+\kappa^2)\omega_0^2}\right)-\frac{1}{2},
\end{eqnarray}
diverges when $\lambda_2\to0$, whereas it vanishes $\langle d_2^\dagger d_2\rangle=0$ in the Dicke model $\lambda_2=0$ because the $b_2$ mode is decoupled from the rest of the system (Fig.~\ref{fig:open_fluc_limits}). Similarly, when $\kappa=0$, the number of independent equations is reduced from 21 to 18, and we have confirmed numerically that the closed-system fluctuations Eq.~\eqref{eq:closed_fluctuations} lies in the solution space. The mathematical singularity manifests itself most apparently when $\lambda_2\le\lambda_1<\frac{1}{2}\sqrt{\omega\omega_0}$, where we find that $\langle d_2^\dagger d_2\rangle$ converges to a negative value in the limit $\kappa\to0$, but has a positive value in the closed system according to Eq.~\eqref{eq:closed_fluctuations} (see Fig.~\ref{fig:open_fluc_limits}).

The cavity and atomic fluctuations for exemplary normal, superradiant and inverted states are solved numerically and shown in Fig.~\ref{fig:open_spectrum_fluc}(c,f,i,l). The cavity fluctuations diverge with a critical exponent of 1 in the vicinity of the stability boundaries (Fig.~\ref{fig:fluc_exponent}). This can be confirmed analytically by performing a Taylor expansion for the cavity fluctuation as a function of $N_1$ or $\theta$ in the vicinity of the stability boundary for the inverted state [Eq.~(7) of the main text].
In Fig.~\ref{fig:non_equal_coupling_spectra}, we show the cavity fluctuations and the imaginary part of the rapidity (i.e., excitation spectrum) of the inverted states for unequal coupling $\lambda_1\neq\lambda_2$. As discussed in the main text, due to the absence of the $\mathrm{U}(1)$ symmetry, the inverted state can be uniquely pinpointed up to a sign in $\theta$ when both cavity fluctuations and excitation spectrum are measured.

The fluctuations are also closely related to the stability of the system. The system is stable when both the cavity and atomic fluctuations are positive, which gives a consistent stability boundary as predicted by the third quantization analysis, providing us with a simpler way to determine the analytical expressions for some stability boundaries. Examples include the stability boundary for the superradiant state [Eq.~(5) of the main text] and the inverted states [Eq.~(6) of the main text].
Notably, in the limit $\kappa\to 0$, neither of these two boundaries converges to their closed-system counterparts, i.e., $\lambda_c=\sqrt{\omega\omega_0}/2$ for the superradiant stability boundary and Eq.~\eqref{eq:inverted_stable_boundary_closed} for the inverted state stability boundary.
To conclude the section, we briefly comment on the geometric interpretation of the stability boundary for the inverted states [Eq.~(6) of the main text]. Written in terms of the three axes $x\equiv\Re\langle \Sigma_{12}\rangle$, $y\equiv\Im\langle \Sigma_{12}\rangle$ and $z\equiv\frac{1}{2}(\langle \Sigma_{22}\rangle-\langle \Sigma_{11}\rangle)$ of the Bloch sphere, the stability boundary is given by
\begin{eqnarray}
	2\lambda_1\lambda_2 x + 2\Omega(\lambda_1^2-\lambda_2^2)z = N\Omega(\lambda_1^2+\lambda_2^2),
\end{eqnarray}
which describes a spherical cap on the Bloch sphere. This geometric interpretation can facilitate the calculation of the area of the stability region.

\section{Dynamical evolution in the open system}

The unusual stability behavior seen in the open system can be confirmed by dynamical evolutions of the system. In this section we introduce the method used to numerically simulate the underlying equations of motion, and discuss the effect induced by the randomness in the initial state.
We simulate the system dynamics based on the Hamiltonian in Eq.~\eqref{eq:hamil_suppmat}. The dynamical behaviors of the system can be well represented by the evolution of the expectation values 
$\langle  a\rangle$, $\langle \Sigma_{01}\rangle$, $\langle \Sigma_{02}\rangle$, $\langle \Sigma_{12}\rangle$, $\langle \Sigma_{00}\rangle$, $\langle \Sigma_{11}\rangle$, $\langle \Sigma_{22}\rangle$.
After considering a mean-field decoupling between the cavity and atomic fields $\langle a \Sigma_{\mu\nu}\rangle\approx \langle a \rangle\langle \Sigma_{\mu\nu}\rangle$, the equations of motion for all these expectation values are given by
\begin{eqnarray}\label{eq:eom_in_simulations}
	\begin{split}
	\frac{d}{dt} \langle a\rangle &= -i\left[(\omega-i\kappa) \langle a\rangle -i\frac{\lambda_1}{\sqrt{N}} (\langle \Sigma_{01}\rangle+\langle \Sigma_{10}\rangle ) + i\frac{\lambda_2}{\sqrt{N}} (\langle \Sigma_{02}\rangle -\langle \Sigma_{20}\rangle)\right], \\
	\frac{d}{dt} \langle \Sigma_{01}\rangle &= -i \left[\omega_0 \langle \Sigma_{01}\rangle + i \frac{\lambda_1}{\sqrt{N}}(\langle a\rangle -\langle a^\dagger \rangle)(\langle \Sigma_{00}\rangle -\langle \Sigma_{11}\rangle) +i\frac{\lambda_2}{\sqrt{N}}(\langle a\rangle +\langle a^\dagger \rangle)\langle \Sigma_{21}\rangle\right], \\
	\frac{d}{dt} \langle \Sigma_{02}\rangle &= -i \left[\omega_0 \langle \Sigma_{02}\rangle - i \frac{\lambda_1}{\sqrt{N}}(\langle a\rangle -\langle a^\dagger \rangle)\langle \Sigma_{12}\rangle - i\frac{\lambda_2}{\sqrt{N}}(\langle a\rangle +\langle a^\dagger \rangle)(\langle \Sigma_{00}\rangle -\langle \Sigma_{22}\rangle)\right], \\
	\frac{d}{dt} \langle \Sigma_{00}\rangle &=  -i\left[i\frac{\lambda_1}{\sqrt{N}}(\langle a\rangle -\langle a^\dagger \rangle)(\langle \Sigma_{01}\rangle-\langle \Sigma_{10}\rangle ) + i\frac{\lambda_2}{\sqrt{N}} (\langle a\rangle +\langle a^\dagger \rangle)(\langle \Sigma_{02}\rangle+\langle \Sigma_{20}\rangle ) \right],  \\
	\frac{d}{dt} \langle \Sigma_{11}\rangle &= +i\left[i\frac{\lambda_1}{\sqrt{N}}(\langle a\rangle -\langle a^\dagger \rangle)(\langle \Sigma_{01}\rangle-\langle \Sigma_{10}\rangle )\right],\\
	\frac{d}{dt} \langle \Sigma_{22}\rangle &= +i\left[ i\frac{\lambda_2}{\sqrt{N}} (\langle a\rangle + \langle a^\dagger \rangle)(\langle \Sigma_{02}\rangle+\langle \Sigma_{20}\rangle ) \right], \\
	\frac{d}{dt} \langle \Sigma_{12}\rangle &= -i\left[-i\frac{\lambda_1}{\sqrt{N}}(\langle a\rangle -\langle a^\dagger \rangle)\langle \Sigma_{02}\rangle - i\frac{\lambda_2}{\sqrt{N}} (\langle a\rangle +\langle a^\dagger \rangle) \langle \Sigma_{10}\rangle\right].
	\end{split}
\end{eqnarray}
We note that the expectation values obey the relation $\langle \Sigma_{\mu\nu}\rangle=\langle \Sigma_{\nu\mu}\rangle^*$. In all simulations below, the total number of particles Eq.~\eqref{eq:total_number} and the two Casimirs Eqs.~\eqref{eq:Casimir} are indeed conserved.

For the simulations of all dynamical evolutions, we prepare a normal state ($\langle \Sigma_{00}\rangle = N$, $\langle \Sigma_{0\nu}\rangle = \langle \Sigma_{\mu\nu}\rangle = 0$ for $\mu,\nu\in\{1,2\}$) with a small initial cavity field $\langle a \rangle = 0.01$. 
The state is then propagated using the equations of motion above and different ramping protocols.
To investigate the multistability of the inverted states [cf. Fig.~3 of the main text], we use the following three ramping protocols:
\begin{subequations}\label{eq:protocol_group1}
	\begin{eqnarray}
		\text{(i)} &\quad& 
		\lambda_1(t) = 	\lambda_2(t) = \begin{cases} \tilde{\omega}^2t/2500,\, &t\le5000/\tilde{\omega} \\ 2\tilde{\omega},\,& t>5000/\tilde{\omega} \end{cases}, \\
		\text{(ii)} &\quad&
		\lambda_1(t) = 	\lambda_2(t) = \begin{cases} \tilde{\omega}^2t/250,\, &t\le500/\tilde{\omega} \\ 2\tilde{\omega},\, &t>500/\tilde{\omega} \end{cases}, \\
		\text{(iii)} &\quad&
		\lambda_1(t) = 	 \begin{cases} \tilde{\omega}^2t/2500,\, &t\le5000/\tilde{\omega} \\ 2\tilde{\omega},\, &t>5000/\tilde{\omega} \end{cases},
		\quad
		\lambda_2(t) = 	 \begin{cases} \tilde{\omega}^2t/2700,\, &t\le5400/\tilde{\omega} \\ 2\tilde{\omega},\,& t>5400/\tilde{\omega} \end{cases};
	\end{eqnarray}
\end{subequations}
while to investigate the stability of the superradiant state, we use the following three ramping protocols:
\begin{subequations}
	\begin{eqnarray}
		\text{(i)} &\quad& 
		\lambda_1(t) = 	1.2\lambda_2(t) = \begin{cases} \tilde{\omega}^2t/14000,\, &t\le14000/\tilde{\omega} \\ \tilde{\omega},\, &t>14000/\tilde{\omega} \end{cases}, \label{eq:ramp_protocol_2_1}\\
		\text{(ii)} &\quad&
		\lambda_1(t) = 	1.2\lambda_2(t) = \begin{cases} \tilde{\omega}^2t/4000,\, &t\le4000/\tilde{\omega} \\ \tilde{\omega},\, &t>4000/\tilde{\omega} \end{cases}, \\
		\text{(iii)} &\quad&
		\lambda_1(t) = 	1.2\lambda_2(t) = \begin{cases} \tilde{\omega}^2t/2000,\, &t\le2000/\tilde{\omega} \\ \tilde{\omega},\, &t>2000/\tilde{\omega} \end{cases}.
	\end{eqnarray}
\end{subequations}

\begin{figure}
	\centering
	\includegraphics[width=\columnwidth]{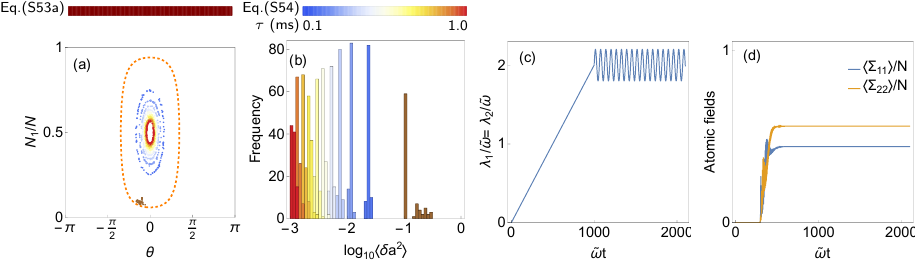}
	\caption{
		(a) The final inverted states in the $N_1$-$\theta$ parameter space and (b) the statistics of their cavity fluctuations of dynamical evolutions starting from different random initial normal states and with different ramp protocols. In panels (a) and (b), the results with protocol Eq.~\eqref{eq:ramp_protocol_3} are shown as blue, yellow and red points/bins, whereas the results with protocol Eq.~\eqref{eq:ramp_protocol_2_1} are shown as brown points/bins. (c) The ramp protocol to test the stability of the inverted states against perturbation in system parameters. (d) The resulting evolutions of the atomic fields. The system remains stable against the perturbation.
	}
	\label{fig:dyn_stability}
\end{figure}

In the main text, we have discussed the sensitivity of the dynamical evolutions on the ramp rate and path. In contrast, the dynamical evolutions show a weak susceptibility to perturbations in the initial condition. 
We initialize the system in the normal state, and impose a random initial condition for the cavity field $\langle a\rangle = 0.01\sqrt{N}(\alpha_\mathrm{Re}+i\alpha_\mathrm{Im})$, where $\alpha_\mathrm{Re}$ and $\alpha_\mathrm{Im}$ are two independent random variables distributed uniformly in the interval $[-1,1]$.  We then numerically evolve the system under the ramp protocol
\begin{eqnarray}\label{eq:ramp_protocol_3}
		\lambda_1(t) = 	\lambda_2(t) = \begin{cases} 2\tilde{\omega}^2t/\tau,\, &t\le \tau\\ 2\tilde{\omega},\,& t>\tau \end{cases}.
\end{eqnarray}
where $\tau$ is the ramp time.
For different ramp times $\tau$ and different initial conditions, the final inverted states the evolutions converge to  are shown in Fig.~\ref{fig:dyn_stability}(a) as red, yellow and blue points. For the same ramp time, the evolutions with different initial normal states converge to different inverted states, all of which lie on the same contour on the $N_1$\nobreakdash-$\theta$ parameter space, and share approximately the same cavity fluctuations [see red, yellow and blue bins in Fig.~\ref{fig:dyn_stability}(b)].
The degeneracy of these final inverted states in dynamical evolutions is a reminiscent of the $\mathrm{U}(1)$ symmetry of the Hamiltonian Eq.~\eqref{eq:hamil_suppmat} for equal couplings, which is also reflected in the mapping Eq.~\eqref{eq:U1inHP}.
Consistently, when we explicitly break the $\mathrm{U}(1)$ symmetry by using, e.g., the ramp protocol Eq.~\eqref{eq:ramp_protocol_2_1}, the dynamical evolutions converge to the vicinity of a single state on the $N_1$\nobreakdash-$\theta$ parameter space, as shown  by the brown points in Fig.~\ref{fig:dyn_stability}(a).  

Finally, we now also test the stability of the inverted states against perturbations in system parameters. For this purpose, we prepare the system in the inverted state and then drive the couplings periodically with ramp protocol shown in Fig.~\ref{fig:dyn_stability}(c). The simulated evolutions of the atomic fields is shown in Fig.~\ref{fig:dyn_stability}(d). Indeed, the system remains stable against such perturbation, which confirms the stability of the inverted states.

\section{Role of atomic symmetry on multistability}\label{sec:sym_multistability}

\subsection{Analysis of systems with $\mathrm{SU}(2)$ atomic symmetry}\label{sec:sym_multistability_SU2}

Here, we argue that  cavity-atom systems  with underlying $\mathrm{SU}(2)$ atomic symmetry, where the cavity couples linearly to the atoms,  cannot host a continuous family of multistable states. We note that linear coupling between cavity and atoms naturally arises as the leading contribution when the cavity light field induces an electric dipole moment in the atomic ensemble~\cite{lekien13,landini18}. We consider a general Hamiltonian describing such a system:
\begin{eqnarray}
	H = \omega a^\dagger a + \omega_0 J_z + \sum_{\mu=x,y,z} \frac{1}{\sqrt{N}}(\lambda_\mu a + \lambda_\mu^* a^\dagger) J_\mu,
\end{eqnarray}
with $J_x$, $J_y$ and $J_z$ the $\mathrm{SU}(2)$ generators.
The mean-field physics of this system is in principle described by four independent expectation values, $\langle a\rangle $, $\langle a\rangle^*$, $\langle J_x\rangle $, and $\langle J_y\rangle$. We emphasize that there are in total four atomic expectation values $\langle J_x\rangle$, $\langle J_y\rangle$, $\langle J_z\rangle$ and $\langle J_0\rangle$, with $J_0$ the identity operator, which are subjected to two constraints. One constraint fixing $\langle J_0\rangle=N/2$ comes from the tracelessness of $\mathrm{SU}(2)$ generators, and corresponding to the conservation of total particles; while the other constraint $\langle J_x
\rangle ^2+\langle J_y\rangle ^2+\langle J_z\rangle ^2=N^2/4$ corresponds to the Casimir of the $\mathrm{SU}(2)$ symmetry~\cite{emary03}.

The mean-field equations of the dissipative system are given by:
\begin{eqnarray}\label{eq:eom_multistability}
	\begin{split}
		\frac{d}{dt} \langle a \rangle &= -i \left[(\omega - i\kappa)\langle a\rangle + \sum_{\mu=x,y,z}\frac{\lambda_\mu^*}{\sqrt{N}} \langle J_\mu\rangle \right] = 0 \\
		\frac{d}{dt} \langle J_x \rangle &= -i \left[-i\omega_0 \langle J_y\rangle + i\frac{1}{\sqrt{N}} (\lambda_y \langle a\rangle + \lambda_y^* \langle a\rangle^*) \langle J_z\rangle - i \frac{1}{\sqrt{N}}(\lambda_z \langle a\rangle + \lambda_z^* \langle a\rangle^*)  \langle J_y\rangle\right] = 0 \\
		\frac{d}{dt} \langle J_y \rangle &= -i \left[-i\omega_0 \langle J_x\rangle - i \frac{1}{\sqrt{N}}(\lambda_x \langle a\rangle + \lambda_x^* \langle a\rangle^*) \langle J_z\rangle + i \frac{1}{\sqrt{N}}(\lambda_z \langle a\rangle + \lambda_z^* \langle a\rangle^*) \langle J_x\rangle\right] = 0.
	\end{split}
\end{eqnarray}
with at least one of $\lambda_\mu$ being nonzero.
To obtain the fixed point states, we set the above equations of motion to zero. There are in total four independent equations in Eqs.~\eqref{eq:eom_multistability}, when we also consider the complex conjugate of the equation for $\langle a\rangle$. To understand the fixed-point solutions of Eqs.~\eqref{eq:eom_multistability}, we enumerate the different potential scenarios below:
\begin{enumerate}
	\item Suppose the fixed-point state has $\langle a\rangle\neq 0$, then the four equations in Eqs.~\eqref{eq:eom_multistability} determine the four parameters. We need to further check for the stability of these steady states. The number of such fixed-point states is finite, and determined by the order of the equations.
	\item Suppose the fixed-point state has $\langle a \rangle = 0$, but the system is nondegenerate $\omega_0\neq0$. We will find $\langle J_x\rangle = \langle J_y\rangle =0$ and $\langle J_z\rangle=\pm N/2$. One of these two fixed-point states could possibly be a dark state when $\lambda_x=\pm i\lambda_y$ and $\lambda_z=0$.
	\item Suppose the fixed-point state has $\langle a \rangle = 0$, and the system is degenerate $\omega_0 = 0$. In this case, the equations for $d\langle J_x\rangle/dt=0$ and $d\langle J_y\rangle/dt=0$ become trivial, whereas the equation for $d\langle a\rangle/dt=0$ is simplified as 
	\begin{eqnarray}
		\sum_{\mu=x,y,z} \Re\lambda_\mu^* \langle J_\mu\rangle = 0,\quad \sum_{\mu=x,y,z} \Im\lambda_\mu^* \langle J_\mu\rangle = 0
	\end{eqnarray}
	\begin{enumerate}
		\item [(a)] When the two equations above are linearly independent, the equations determine a single fixed-point state. This is the case when there are at least two nonzero $\lambda_\mu$ not having the same argument, or in other words, when there does not exist a gauge transformation $a\to ae^{i\phi}$ such that all nonzero $\lambda_\mu$ become real.
		Such a fixed-point state is stable, and it is a dark state satisfying $a|\Psi\rangle=0$, $H|\Psi\rangle=0$.
		\item [(b)] When the criterion above is not satisfied, the two equations become linearly dependent. There is only one nontrivial equation, determining a circle on the Bloch sphere. We notice that after the abovementioned transformation $a\to ae^{i\phi}$, all coupling strengths $\lambda_\mu$ become real, and the Hamiltonian reads
		\begin{subequations}
			\begin{eqnarray}
				H = \omega a^\dagger a + (a+a^\dagger) \sum_{\mu=x,y,z} \frac{\lambda_\mu}{\sqrt{N}} J_\mu.
			\end{eqnarray}
			A further transformation in the atomic operator gives
			\begin{eqnarray}
				H = \omega a^\dagger a + \frac{\lambda}{\sqrt{N}}(a+a^\dagger) \tilde{J}_z.
			\end{eqnarray}
		\end{subequations}
		We have thus recovered a Dicke model with vanishing atomic frequency, which transitions to the superradiant state at $\lambda=0$~\cite{emary03}. The fixed-point states of this Hamiltonian are thus all unstable as soon as $\lambda$ is switched on.
\end{enumerate}

\end{enumerate}

In conclusion, we have shown that a cavity-atom system with an $\mathrm{SU}(2)$ atomic system and a linear coupling between cavity and atoms always has a finite number of steady states, among which at most one dark state. This argument is crucially based on a comparison between the number of unknown parameters and the number of equations governing them. There is only one case, i.e. case 3(b), where the equations are underdetermined, leading to a continuous family of fixed-point states, which all turn out to be unstable.
 The arguments can be naturally extended to $\mathrm{SU}(2)$ atomic systems coupled to multiple cavities. We also expect that they are generalizable to cases where additional nonlinear couplings arise on top of the linear couplings, because these nonlinear couplings will only induce higher-order effects compared to the linear couplings when $\langle a\rangle=0$.

\subsection{Analysis of systems with $\mathrm{SU}(3)$ atomic symmetry}

Compared to $\mathrm{SU}(2)$ systems, the description of the $\mathrm{SU}(3)$ atomic Hilbert space requires more expectation values, potentially leading to the realization of multistable nearly dark states. For simplicity, we take as example a Hamiltonian similar to Eq.~\eqref{eq:hamil_suppmat} of our system:
\begin{eqnarray}\label{eq:hamil_non_degenerate}
	H = \omega a^\dagger a + \omega_1 \Sigma_{11} + \omega_2\Sigma_{22}
	+ \frac{i\lambda_1}{\sqrt{N}}(a - a^\dagger)(\Sigma_{01}+\Sigma_{10}) 
	+ \frac{i\lambda_2}{\sqrt{N}}(a^\dagger + a)(\Sigma_{02}-\Sigma_{20}).
\end{eqnarray}
where the three levels are now possibly nondegenerate.

We first discuss the total number of independent expectation values. There are two cavity expectation values $\langle a\rangle$, $\langle a\rangle^*$ and nine atomic expectation values $\langle \Sigma_{\mu\nu}\rangle$ with $\mu,\nu\in\{0,1,2\}$. Similar to the case of $\mathrm{SU}(2)$, there is a constraint Eq.~\eqref{eq:total_number} coming from $\mathrm{SU}(3)$ generators being traceless, and there are two additional constraints Eqs.~\eqref{eq:Casimir} corresponding to the two Casimirs of the $\mathrm{SU}(3)$ symmetry. There are thus eight independent cavity and atomic expectation values in total.

We then discuss the total number of independent equations. Based on the previous discussion of the $\mathrm{SU}(2)$ system, one would typically expect the situation where the number of parameters is larger than the number of equations relating them, when the atomic levels are partially degenerate and the cavity field is zero. Indeed, when the cavity field is vanishing in the mean-field limit $\langle a\rangle=0$, the mean-field equations of motion governing the expectation values are given by
\begin{eqnarray}
	\begin{split}
		\frac{d}{dt}\langle a\rangle &= -\frac{2\lambda_1}{\sqrt{N}}\Re\langle\Sigma_{01}\rangle + \frac{2i\lambda_2}{\sqrt{N}}\Im\langle\Sigma_{02}\rangle = 0 \\
		\frac{d}{dt}\langle \Sigma_{01}\rangle &= -i\omega_1 \langle \Sigma_{01}\rangle = 0 \\
		\frac{d}{dt}\langle \Sigma_{02}\rangle &= -i\omega_2 \langle \Sigma_{02}\rangle = 0 \\
		\frac{d}{dt}\langle \Sigma_{12}\rangle &= -i(\omega_2-\omega_1) \langle \Sigma_{12}\rangle =0.
	\end{split}
\end{eqnarray}
Together with their respective complex conjugated version, there are eight equations in total. When the levels become degenerate, some of them could become trivial, making the set of equations underdetermined, and potentially leading to multistable steady states.

When $\omega_1\neq0$, $\omega_2\neq0$ and $\omega_1\neq\omega_2$, and thus all equations are nontrivial, the eight unknown expectation values can be completely and uniquely determined. There are thus a finite number of steady states with $\langle a\rangle=0$. On the contrary, in the degenerate limit $\omega_1=\omega_2\neq0$ where Eq.~\eqref{eq:hamil_suppmat} is recovered, we find two of the equations $d\langle \Sigma_{12}\rangle/dt=0$ and $d\langle \Sigma_{12}\rangle^*/dt=0$ become trivial, resulting in an underdetermined set of equations and potentially giving rise to a two-parameter family of multistable states. We emphasize that in principle a stability analysis on these fixed-point states are further required to confirm their stability.
Finally, when all three levels are degenerate $\omega_1=\omega_2=0$, the system could even potentially host an even larger six-parameter family of multistable steady states.

With this example we have illustrated that a  larger symmetry of the Hilbert space indeed  permits the realization of a continuous family of steady states.

\section{Modifications to the model and their consequences}

Many results presented in our work rely crucially on the specific structure of our Hamiltonian Eq.~\eqref{eq:hamil_suppmat}. In this section, we discuss the preliminary results for a few modifications to the model and their consequences.

\subsection{A $\Lambda$-shaped system with negative $\omega_0$}

We first discuss the situation where $\omega_0$ changes sign and the V-shaped atomic system becomes $\Lambda$-shaped.
With $\omega_0<0$, the ground state in the closed system has now macroscopic occupation in levels $|1\rangle$ and $|2\rangle$ for couplings smaller than the Dicke critical coupling $\lambda_1<\lambda_c$, $\lambda_2<\lambda_c$, since their energies are lower than the energy of the $|0\rangle$ level.
In the open system, we find that the $|0\rangle$ state is always unstable unless the coupling to light is completely turned off $\lambda_1=\lambda_2=0$. On the other hand, a subset of the states with macroscopic occupation in levels $|1\rangle$ and $|2\rangle$ remains stable. Although the stability boundary in the $N_1$\nobreakdash-$\theta$ phase space is now different from Eq.~(6) of the main text, the dark state $|D\rangle$ is always stable. We also note that the stability of the superradiant states is not affected. Importantly, we remark that for $\omega_0<0$, the dissipation no longer induces a population inversion in the atoms.

\subsection{A system with nondegenerate high-energy levels}

Many nontrivial phenomena in the open system depends crucially on the degeneracy of the two high-energy levels, $|1\rangle$ and $|2\rangle$. We now study a Hamiltonian with this degeneracy lifted as described in Eq.~\eqref{eq:hamil_non_degenerate}.
As discussed in Sec.~\ref{sec:sym_multistability}, as soon as the degeneracy is lifted, the continuous family of steady states stops to exist, giving rise to a finite number of steady states. In this specific system, the dark state also vanishes. This is in great contrast to the case where the levels are coupled through classical light, and the dark state exists for both degenerate and nondegenerate levels~\cite{bergmann98}.  
Therefore, with a tiny energy splitting $\Delta\omega=|\omega_1-\omega_2|$, there is a large region in the phase diagram where no steady state exists, and the system shows oscillatory behaviors.
However, as the energy splitting increases, we observe that the normal state becomes stable again in a larger and larger part of the phase diagram. 
As a result, when the energy splitting  becomes comparable to the energies themselves ${\Delta\omega \sim \omega_1}$, we recover physics very reminiscent of a Dicke-like model, wherein  the normal state becomes stable again at low coupling in the open system.  An indepth study of the nondegenerate case merits a separate work.
As results, the phase diagram now consists of a normal phase, two separate superradiant phases, and an oscillatory region.
With these preliminary results, we expect more interesting phenomena to be discovered in this nondegenerate scenario.

\bibliographystyle{apsrev}
\bibliography{References}